\documentclass[preprint,showpacs,preprintnumbers,amsmath,amssymb,floatfix]{revtex4-1}
\newcommand{\newwidth}{0.675\textwidth}
\newcommand{\newheight}{0.45\textwidth}

\newcommand{\newwidthprime}{0.225\textwidth}
\newcommand{\newheightprime}{0.30\textwidth}
\usepackage{graphicx}
\usepackage{dcolumn,psfrag}
\usepackage{bm,verbatim}
\usepackage{color}

\begin{document}

\title{Effect of broadening in the weak coupling limit of vibrationally coupled electron transport 
through molecular junctions and the analogy to quantum dot circuit QED systems} 

\author{Rainer H\"artle$^{1}$}
\author{Manas Kulkarni$^{2}$}
\affiliation{
$^1$ Institut f\"ur theoretische Physik, Georg-August-Universit\"at G\"ottingen, Friedrich-Hund-Platz 1, D-37077 G\"ottingen, Germany. \\
$^2$ Department of Physics, New York City College of
 Technology, The City University of New York, Brooklyn, NY 11201, USA.
}

\date{\today}

\begin{abstract}
We investigate the nonequilibrium population of a vibrational mode in the steady-state of 
a biased molecular junction, using a rate equation approach. We focus on the limit of weak 
electronic-vibrational coupling and show that, in the resonant transport regime and for sufficiently 
high bias voltages, the level of vibrational excitation increases with decreasing coupling strength, 
assuming a finite and non-zero value. An analytic behavior with respect to the electronic-vibrational 
coupling strength is only observed if the influence of 
environmental degrees of freedom is explicitly taken into account. 
We consider the influence of three different types of broadening: 
hybridization with the electrodes, thermal fluctuations and the coupling to a thermal heat bath. 
Our results apply to vibrationally coupled electron transport 
through molecular junctions but also to quantum dots coupled to a microwave cavity, 
where the photon number can be expected to exhibit a similar behavior.  
\end{abstract}

\pacs{85.35.-p, 73.63.-b, 72.10.Di}

\maketitle

\section{Introduction}

The interplay of electronic degrees of freedom with phonons or photons is 
relevant for a wide range of condensed matter systems. This includes systems 
where electrons interact with local vibrational degrees of freedom such as, for example, in molecules, 
or with delocalized lattice vibrations (phonons) such as, for example, in solids. 
The interaction of electrons and photons 
can be understood on similar grounds, identifying the electric field strength with the displacement 
of a vibrational or a phonon degree of freedom. Thus, for example, matter-light interaction is often described 
by Hamiltonians \cite{SchaeferWegener,Delbecq2011} that are very similar to the ones used to describe 
electron-phonon interactions \cite{Mahan81}.

The coupling results in mutual excitation and deexcitation processes. 
For example, phonons or photons can be excited via a deexcitation of the electronic subsystem. 
Such an excitation decays typically fast, 
if the phonon or photon field is delocalized and the excitation energy is transferred efficiently 
away from the place of interaction. This is the situation when the electrons 
are under the influence of an external electric field 
or interact with phonons in a solid. The situation is different 
for nanostructures where the bosonic degrees of freedom are localized. 
The simplest realization of such a scenario is a molecule. Photons can be localized inside a 
cavity \cite{Raimond2001,Wallraff2004,Chiorescu2004}. 
In both cases, the bosonic degrees of freedom can be driven into highly excited 
nonequilibrium states \cite{Semmelhack,Hartle09,Hartle2010b,Romano10,Schiro2014,Abdullah2014,Kulkarni2014,Schinabeck2014}. 
It is challenging to understand these states, because they cannot be described via a 
Bose-Einstein distribution function. Nevertheless, these states 
determine the functionality and efficiency of many devices 
that may be used for nanoelectronic applications 
\cite{Schoeller01,Cuniberti05,Koch05b,Zazunov06,Hartle09,Saito2009,Leijnse09,Romano10,cuevasscheer2010,Hartle2010b,ChenWang2012,Hartle2013}, 
quantum information processing \cite{Skolnick2004,Cuniberti05,Schoelkopf2008,Juska2014}, 
solar energy conversion \cite{Selinsky2013,ChenWang2014,Wu2014,Ajisaka2015} 
or the control of chemical processes \cite{Brumer03,Pascual03,Hartle2010,Kosov2011,Volkovich2011b}.

In this work, we investigate the nonequilibrium properties of a vibrational mode 
in a single-molecule junction. Thereby, the molecule is contacted by two electrodes which represent 
macroscopic electron reservoirs. This can be achieved, for example, using a scanning tunneling 
microscope, electromigration or break-junction techniques \cite{Nitzan01,Tao2006,Selzer06,Chen07,cuevasscheer2010}. 
A bias voltage can be applied such that electrons tunnel through the molecule. 
Due to the small size and mass of a molecule, 
the tunneling electrons are likely to interact with the molecules vibrational modes, leading 
to inelastic tunneling processes where vibrational motion is excited and deexcited 
\cite{May02,Flensberg03,Mitra04,WuNazinHo04,Galperin04,Frederiksen04,Cizek04,Wegewijs05,Galperin07,Avriller2009,Schmidt2009,Haupt2009,Hartle09,Leijnse09,Fehske2010,Urban2010,Wohlman2010,Esposito2010,Godby2012,Hartle2013,Schinabeck2014}. 
This behavior has been observed in a number of experiments \cite{Natelson04,Kushmerick04,Qiu04,Sapmaz05,Pasupathy05,
Sapmaz06,Thijssen06,Parks07,Boehler07,Leon2008,Repp2009,Huettel2009,Tao2010,Ballmann2010}. 
The observation, however, is often indirect, because a direct measurement of the level of vibrational 
excitation, for example via Raman spectroscopy \cite{Natelson2008,Ioffe08,Natelson2011}, 
is very involved. 

\begin{figure}
\includegraphics[width=7.2cm]{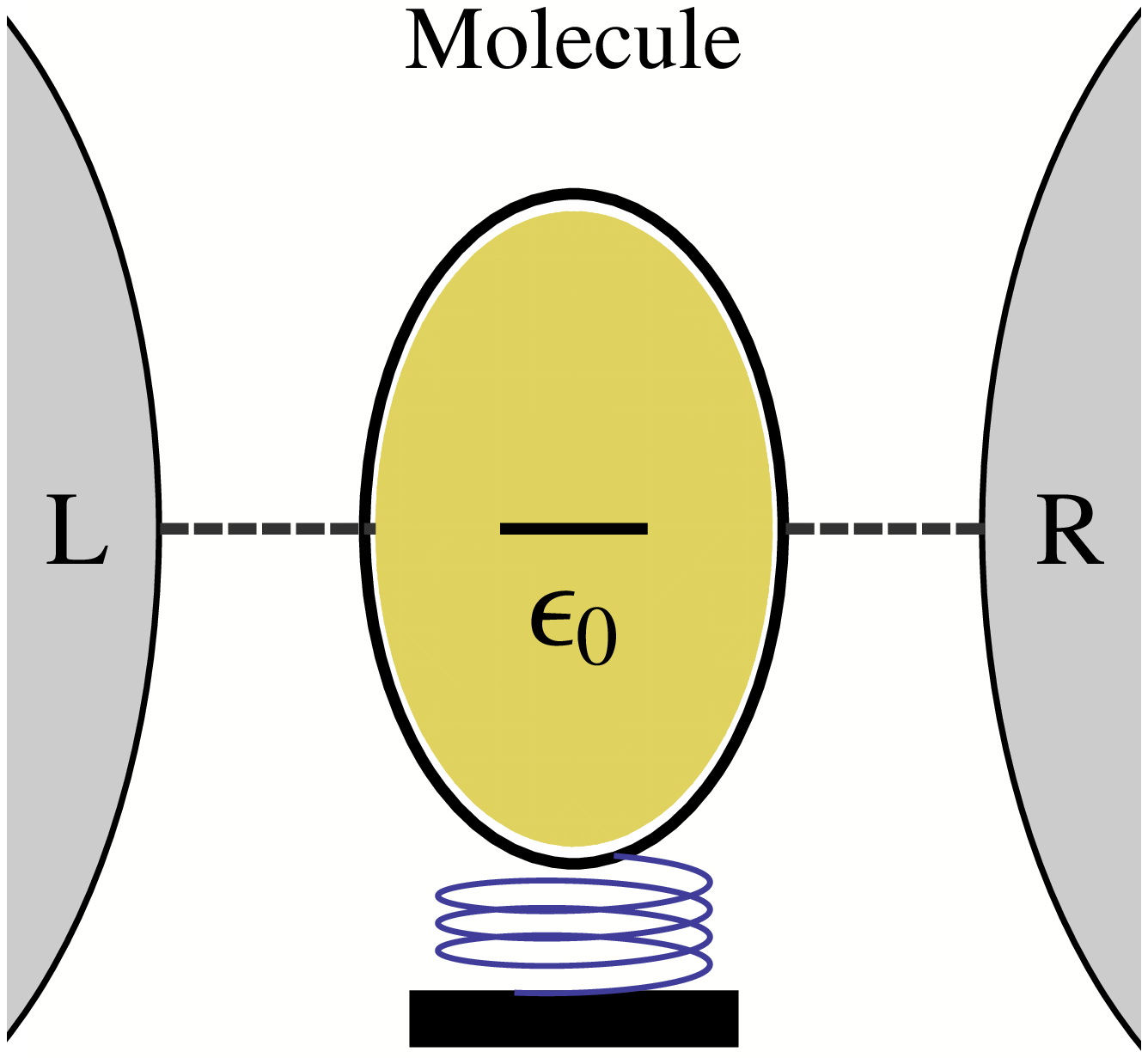}
\qquad\quad
\includegraphics[width=7.2cm]{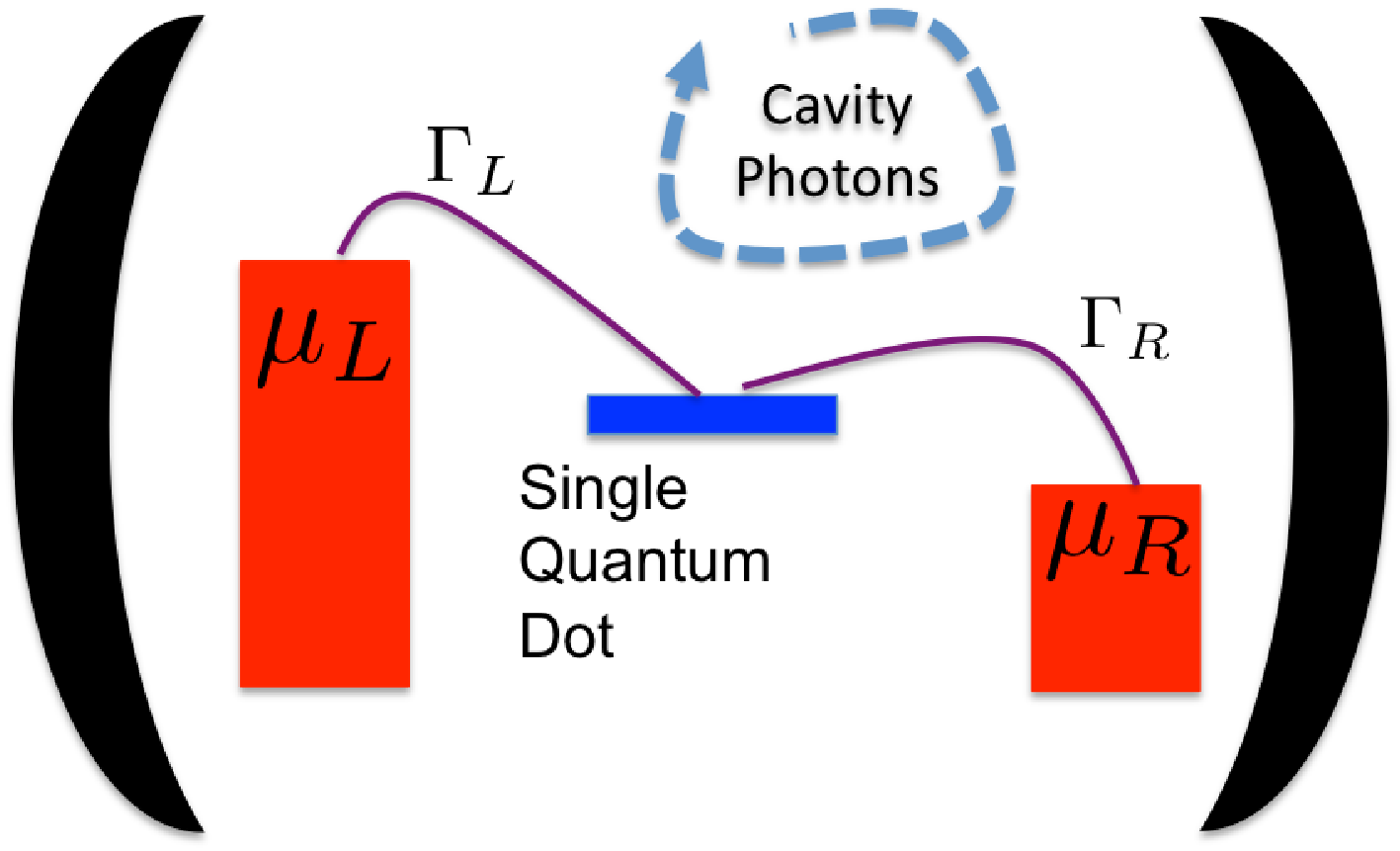}
\caption{\label{hagt} Schematic figure of a single-molecule junction (left) and a single quantum dot circuit-QED setup (right). 
The coupling between a local vibrational mode of the molecule, which is depicted by a spring, to one of the molecular 
electronic levels can be described in the same \emph{Holstein-like} way as the coupling between 
the microwave photons of a cavity and the electronic levels of a quantum dot inside the cavity. }
\end{figure}

The level of vibrational excitation may be more easily and more reliably accessed in quantum dot systems 
which are coupled to a microwave cavity (circuit quantum electrodynamics) 
\cite{Yoshie2004,Reithmaier2004,Petersson2011,Delbecq2011,Delbecq2013,Liu2014}. 
Here, the quantum dots take on a similar role 
as the molecule in a molecular junction, while the number of photons inside the cavity correspond to the excitation 
number of a local vibrational mode. Infact, measurements of the photon number have been reported 
recently \cite{Viennot2014}. 
Identifying again the electric field strength with the displacement coordinate of a vibrational mode, 
the theoretical description of these systems is very similar 
\cite{Delbecq2011,Delbecq2013,Schiro2014,Abdullah2014,Kulkarni2014,Cottet2015} (cf.\ Fig.\ \ref{hagt}). 
In contrast to single-molecule junctions, however, where the contact geometry and other 
(\emph{e.g.}\ internal) coupling parameters are hard to control, 
experiments on quantum dot systems can be carried out with greater precision and more handles of control, 
including, for example, the wide tunability of the electron-photon coupling strength \cite{Delbecq2013,Viennot2014}. 
Given these remarkable analogies, vibrational dynamics in a molecular junction can be used to 
understand quantum impurity circuit-QED systems and vice versa.

In this contribution, we focus on the counterintuitive prediction that, in a molecular junction at sufficiently 
high bias voltages, the level of vibrational excitation increases with decreasing electronic-vibrational 
coupling strengths. This phenomenon has been reported in a number of theoretical 
studies \cite{Mitra04,Semmelhack,Avriller2010,Hartle2011,Ankerhold2011} and explained 
via the suppression of electron-hole pair creation processes \cite{Hartle2011}. 
The latter study was based on Born-Markov theory and pointed out the connection between the limits  
of an infinite bias voltage and a vanishing electronic-vibrational coupling strength. Thereby, the hybridization 
of the junctions levels with the electrodes, thermal broadening and the effect of a heat bath has been discarded. 
We take a big step forward and extend these considerations by broadening effects, verifying that the slope of the 
vibrational excitation level as function of electronic-vibrational coupling  
can be negative over a wide range of parameters. 
Due to broadening, the level of vibrational excitation 
stays finite in the limit of a vanishing coupling and does not increase indefinitely 
as the coupling constant vanishes. Moreover, we show that the non-analytic behavior is an artifact
of the model itself. For non-zero coupling strengths and/or in presence of 
a heat bath (analogous to the leakage of photons in the cavity context) 
we find an analytic behavior. 
While the electronic-vibrational coupling is hard to control in molecular junctions, 
quantum dot realizations may give a direct access to this phenomenon, that is a decrease of the 
level of vibrational excitation or photon number with an increasing coupling strength.

The article is organized as follows. In Sec.\ \ref{theory}, we outline the theoretical methodology. 
This includes our model of a molecular junction (Sec.\ \ref{modelHamSec}) and 
the Born-Markov theory that we use to solve this transport problem (Secs.\ \ref{bornmarkovtheory}). 
In addition, we outline an extension of the approach by higher order processes, 
which involves off-resonant pair creation processes and cotunneling in terms of transition rates 
(Sec.\ \ref{ratetheory}). Our results are presented in Sec.\ \ref{results}. 
We first review the basics of the phenomenon (Sec.\ \ref{Basics}) and continue to discuss the 
effect of broadening on the level of vibrational excitation in the limit of vanishing electronic-vibrational 
coupling (Sec.\ \ref{thermBroad}). The next section (Sec.\ \ref{BroadCoTun}) 
is devoted to the effect of excitation and deexcitation processes 
due to inelastic cotunneling \cite{Koch2006,Lueffe} and off-resonant electron-hole pair creation 
processes \cite{Volkovich2011b}. We show that these effects can only be properly accounted for 
if broadening effects are considered, while, otherwise, these processes have no impact on our phenomenon of interest. 
In Sec.\ \ref{heatbathsec}, we will finally address broadening of the vibrational levels 
due to coupling to a heat bath.

\section{Theory}
\label{theory}

\subsection{Model hamiltonian}
\label{modelHamSec}

We consider vibrationally coupled electron transport through a nanostructure that 
can be represented by a single electronic state. The Hamiltonian of this transport 
scenario can be written as follows (using units where $\hbar=1$)
\begin{eqnarray}
H &=& \epsilon_{0} d^{\dagger}d + 
\sum_{k\in\text{L,R}} \epsilon_{k} c_{k}^{\dagger}c_{k} + \sum_{k\in\text{L,R}} ( V_{k} 
c^{\dagger}_{k} d + \text{h.c.} ) + \Omega
a^{\dagger}a + \lambda (a+a^{\dagger}) d^{\dagger}d \\
&& + \sum_{\beta} \omega_{\beta} b_{\beta}^{\dagger}b_{\beta} + \sum_{\beta} \eta_{\beta}  (a+a^{\dagger}) 
(b_{\beta}+b_{\beta}^{\dagger})\nonumber
\end{eqnarray}
Here, $\epsilon_{0}$ denotes the energy of the electronic state. 
It is coupled to electronic states with energies $\epsilon_{k}$ that are located 
in the left (L) and the right (R) electrode. 
The corresponding coupling matrix elements are given by $V_{k}$. 
Each of these states is addressed by creation/annihilation 
operators $d^{\dagger}$/$d$ and $c_{k}^{\dagger}/c_{k}$, respectively. 
For the vibrational degrees of freedom, we employ a simplified picture, where 
we use a single harmonic mode with frequency $\Omega$. 
The corresponding creation and annihilation operators are $a^{\dagger}$ and $a$. 
The coupling strength between the harmonic mode 
and the electronic state is denoted by $\lambda$. 
In addition, we consider a direct coupling of the local vibrational degree of freedom 
to a bath of oscillators \cite{Weiss93,Thoss98,Hartle,Hartle2010}. They are 
addressed by the creation and annihilation operators $b_{\beta}^{\dagger}$ and $b_{\beta}$. 
The corresponding coupling strengths are given by $\eta_{\beta}$. 
Thus, we account for a coupling of the local vibrational mode to the phonon modes in the electrodes \cite{Seidemann10}, 
other environmental degrees of freedom, such as, \emph{e.g.},  
in an electrochemically gated molecular junction \cite{Tao2006} and/or 
intramolecular vibrational energy redistribution \cite{Mukamel1980,Freed,Zewail,Nesbitt1996}. 
In the quantum dot circuit QED context, 
these terms simulate the leakage of photons out of the cavity.

Using the small polaron transformation \cite{Mitra04}, which, in the presence of a heat bath, 
can be carried out iteratively as long as $4\sum_{\beta}  
\frac{\left\vert \eta_{\beta} \right\vert^{2} }{\Omega\omega_{\beta}}<1$ \cite{Galperin06,Hartle,Hartle09}, 
the Hamiltonian $H$ can be pre-diagonalized. The transformed Hamiltonian 
$H\rightarrow\overline{H}$  
is given by 
\begin{eqnarray}
 \overline{H} &=& \overline{H}_{\text{S}} + \overline{H}_{\text{L+R+bath}} + \overline{H}_{\text{SL+SR+Sbath}}, \\
 \overline{H}_{\text{S}} &=& \overline{\epsilon}_{0} d^{\dagger}d  + \Omega a^{\dagger} a,  \\
 \overline{H}_{\text{L+R+bath}} &=& \sum_{k} \epsilon_{k} c_{k}^{\dagger}c_{k} + \sum_{\beta}\omega_{\beta} 
 b_{\beta}^{\dagger}b_{\beta}, \\ 
 \overline{H}_{\text{SL+SR+Sbath}} &=& \sum_{k} ( V_{k} X  c_{k}^{\dagger}d + \text{h.c.} ) + 
 \sum_{\beta} \eta_{\beta}  (a+a^{\dagger}) 
(b_{\beta}+b_{\beta}^{\dagger}). 
\end{eqnarray}
It involves a system (S) Hamiltonian, $\overline{H}_{\text{S}}$, 
which comprises the polaron-shifted electronic level, 
$\overline{\epsilon}_{0}=\epsilon_{0}-\lambda^{2}/\Omega$, 
and the harmonic mode. The leads degrees of freedom and the heat bath are 
included in $\overline{H}_{\text{L+R+bath}}$. 
The term $\overline{H}_{\text{SL+SR+Sbath}}$
describes the coupling between the electronic state and the leads and 
between the local vibrational mode and the heat bath. 
As a result of the polaron transformation, there is no 
direct electronic-vibrational coupling term in $\overline{H}_{\text{S}}$, but 
the coupling matrix elements $V_{k}$ 
become dressed by the shift operator $X=\text{exp}((\lambda/\Omega)(a-a^{\dagger}))$. 
Throughout this work, we assume the wide-band approximation where the 
level-width function 
\begin{eqnarray}
 \Gamma_{K}(\epsilon) &=& 2\pi \sum_{k\in K} \vert V_{k} \vert^{2} \delta(\epsilon-\epsilon_{k}) 
\end{eqnarray}
is approximated by a constant $\Gamma_{K}(\epsilon)=\Gamma_{K}$ ($K\in\{\text{L,R}\}$). 
Moreover, we characterize the effect of the heat bath via the constant dissipation rate 
\begin{eqnarray}
 \gamma_{\text{bath}} &=& 2\pi \sum_{\beta} \vert \eta_{\beta} \vert^{2} \delta(\Omega-\omega_{\beta}).  
\end{eqnarray}
Note that, here and in the following, 
we neglected any renormalization effects due to the coupling to the thermal heat bath 
and assume $\sum_{\beta}  
\frac{\left\vert \eta_{\beta} \right\vert^{2} }{\Omega\omega_{\beta}}\ll1$.

\subsection{Born-Markov master equation approach} 
\label{bornmarkovtheory}

We address the transport problem, which is described by the Hamiltonian $\overline{H}$, 
employing Born-Markov theory. The corresponding master equation is well established 
\cite{May02,Mitra04,Lehmann04,Harbola2006,Volkovich2008,Hartle09,Hartle2010b}. 
The central quantity of this theory is the reduced density matrix $\sigma$. 
It is determined by the equation of motion 
\begin{eqnarray}
\label{genfinalME}
\frac{\partial \sigma(t)}{\partial t} &=& -i \left[ 
\overline{H}_{\text{S}} , \sigma(t) \right] \\
&& - \int_{0}^{\infty} \text{d}\tau\, 
\text{tr}_{\text{L+R+bath}}\lbrace \left[ \overline{H}_{\text{SL+SR+Sbath}} , \left[ \overline{H}_{\text{SL+SR+Sbath}}(\tau), \sigma(t) 
\sigma_{\text{L}} \sigma_{\text{R}} \sigma_{\text{bath}} \right] \right] \rbrace , \nonumber
\end{eqnarray}
where 
\begin{eqnarray}
\overline{H}_{\text{SL+SR+Sbath}}(\tau) &=& 
\text{e}^{-i(\overline{H}_{\text{S}}+\overline{H}_{\text{L+R+bath}})\tau} 
\overline{H}_{\text{SL+SR+Sbath}} \text{e}^{i(\overline{H}_{\text{S}}+\overline{H}_{\text{L+R+bath}})\tau}   
\end{eqnarray}
and $\sigma_{K}$ and $\sigma_{\text{bath}}$ represent 
the equilibrium density matrix of lead $K$ and the heat bath, respectively. 
This equation of motion constitutes a second-order expansion of the 
Nakajima-Zwanzig equation \cite{Nakajima,Zwanzig} with respect 
to the coupling term $\overline{H}_{\text{SL+SR+Sbath}}$, where, in addition, 
the so-called Markov approximation is employed.

We evaluate the master equation (\ref{genfinalME}) with the 
product basis $\vert 0 \rangle\vert \nu \rangle$ and $\vert 1 \rangle\vert \nu \rangle$, 
which refers to the $\nu$th level of the harmonic mode ($\nu\in\mathbb{N}_{0}$) and the 
unoccupied $\vert 0 \rangle$ and occupied electronic level $\vert 1 \rangle$, respectively. 
In the following, we consider only the diagonal elements of the reduced density matrix, because 
the respective off-diagonal elements do not play a role in what follows (cf.\ the discussion 
in the appendix of Ref.\ \cite{Hartle2010b}). 
The diagonal elements can be written as 
\begin{eqnarray}
 \sigma_{0}^{\nu} \equiv \langle 0 \vert \sigma^{\nu} 
\vert 0 \rangle \equiv \langle 0 \vert \langle \nu \vert \sigma \vert \nu \rangle \vert 0 \rangle, \\
 \sigma_{1}^{\nu} \equiv \langle 1 \vert \sigma^{\nu} 
\vert 1 \rangle \equiv \langle 1 \vert \langle \nu \vert \sigma \vert \nu \rangle \vert 1 \rangle.  
\end{eqnarray}
In the steady state regime, where $\partial_{t}\sigma=0$, 
the master equation (\ref{genfinalME}) thus reduces to the rate equations 
\begin{eqnarray}
\label{explicitsecondorderrates}
0 &=& - \sum_{K,\nu'} \Gamma_{K,0\rightarrow1}^{\nu\rightarrow\nu'} \sigma^{\nu}_{0} 
+ \sum_{K,\nu'} \Gamma_{K,1\rightarrow0}^{\nu'\rightarrow\nu} \sigma^{\nu'}_{1} \\
&& + \gamma_{\text{bath}} \left( 1 - \text{e}^{-\frac{\Omega}{k_\text{B}T}} \right)^{-1} \left( 
\nu \sigma^{\nu}_{0} - (\nu+1) \sigma^{\nu+1}_{0} \right) 
+ \gamma_{\text{bath}} \left( \text{e}^{\frac{\Omega}{k_\text{B}T}} - 1 \right)^{-1} \left( 
(\nu+1) \sigma^{\nu}_{0} - \nu \sigma^{\nu-1}_{0} \right)  , \nonumber \\
0 &=& - \sum_{K,\nu'} \Gamma_{K,1\rightarrow0}^{\nu\rightarrow\nu'} \sigma^{\nu}_{1} 
+ \sum_{K,\nu'} \Gamma_{K,0\rightarrow1}^{\nu'\rightarrow\nu} \sigma^{\nu'}_{0}, \nonumber\\   
&& + \gamma_{\text{bath}} \left( 1 - \text{e}^{-\frac{\Omega}{k_\text{B}T}} \right)^{-1} \left( 
\nu \sigma^{\nu}_{1} - (\nu+1) \sigma^{\nu+1}_{1} \right) 
+ \gamma_{\text{bath}} \left( \text{e}^{\frac{\Omega}{k_\text{B}T}} - 1 \right)^{-1} \left( 
(\nu+1) \sigma^{\nu}_{1} - \nu \sigma^{\nu-1}_{1} \right)  , \nonumber 
\end{eqnarray}
with the rates 
\begin{eqnarray}
 \Gamma_{K,0\rightarrow1}^{\nu_{i}\rightarrow\nu_{f}} &=&  f_{K}(\overline{\epsilon}_{0}+\Omega(\nu_{f}-\nu_{i})) 
\Gamma_{K} X_{\nu_{i}\nu_{f}} X_{\nu_{f}\nu_{i}}^{\dagger} , \\
 \Gamma_{K,1\rightarrow0}^{\nu_{i}\rightarrow\nu_{f}} &=&  \left(1-f_{K}(\overline{\epsilon}_{0}+\Omega(\nu_{i}-\nu_{f}))\right) 
\Gamma_{K} X_{\nu_{f}\nu_{i}} X_{\nu_{i}\nu_{f}}^{\dagger}, 
\end{eqnarray}
the transition matrix elements 
\begin{eqnarray}
 X_{\nu\nu'} &=&  \langle \nu \vert X \vert \nu' \rangle,  
\end{eqnarray}
the Fermi distribution function associated with lead $K$
\begin{eqnarray}
 f_{K}(\epsilon) &=& \frac{1}{1+\text{exp}\left( \frac{\epsilon-\mu_{K}}{k_\text{B}T} \right)}  
\end{eqnarray}
and $k_\text{B}$ the Boltzmann constant. 
Thereby, we use the Fermi level of the setup as the zero of energy, $\epsilon_{\text{Fermi}}=0$, 
and a symmetric drop of the bias voltage $\Phi$ at the contacts, \emph{i.e.}\ 
the chemical potentials in the leads are given by $\mu_{\text{L/R}}=\pm e\Phi/2$. Note that 
the specific way of the voltage drop is not going to be decisive for our discussion.

The rate equations (\ref{explicitsecondorderrates}) are simplified in the sense 
that all principal value terms, which would emerge from the 
$\int\text{d}\tau$-integral in Eq.\ (\ref{genfinalME}), are neglected. These 
terms describe the renormalization of the molecular energy levels due to the 
molecule-lead coupling \cite{Harbola2006}. In a recent study of 
charge-transfer dynamics in a double quantum dot system \cite{Hartle2014}, 
it was shown that these terms are particularly important in the presence of quasi-degenerate levels. 
However, since we consider $\Omega\gg\Gamma_{K}$, such renormalization effects can be safely 
neglected for the purpose of our discussion.

\subsection{Cotunneling rates} 
\label{ratetheory}

Due to the second order expansion in $\overline{H}_{\text{SL+SR+Sbath}}$, 
the Born-Markov master equation (\ref{genfinalME}) 
can only describe resonant processes. 
Higher-order expansions of the Nakajima-Zwanzig equation 
include non-resonant processes, like cotunneling \cite{May02,Pedersen05,Koch2006,Leijnse09}. 
These processes involve not only additional transport channels \cite{Lueffe} 
but also off-resonant pair creation processes \cite{Volkovich2011b}. 
Both process classes result in a broadening of energy levels and are, therefore, 
particularly interesting in the present context.

Restricting the discussion to 
the populations $\sigma_{0/1}^{\nu}$, the effect of cotunneling processes 
can be subsumed in transition rates. 
The cotunneling rates for vibrationally coupled electron transport 
have been derived by Koch \emph{et al.}\ \cite{Koch2006}. 
For processes involving the empty state, they read 
\begin{eqnarray}
\label{cotrate}
 \gamma_{K_{i}\rightarrow K_{f},0\rightarrow0}^{\nu_{i}\rightarrow\nu_{f}} 
 &=& \frac{1}{2\pi} \Gamma_{K_{i}} \Gamma_{K_{f}} 
\sum_{\nu} \vert X_{\nu_{f}\nu} X_{\nu_{i}\nu} \vert^{2} J(\mu_{K_{i}}, 
\mu_{K_{f}}-\Omega (\nu_{i} -\nu_{f} ),\overline{\epsilon}_{0}-\Omega(\nu_{i}-\nu))\nonumber\\
 &&\hspace{-2.5cm} + \frac{1}{2\pi} \Gamma_{K_{i}} \Gamma_{K_{f}} 
\sum_{\nu'\neq\nu} X_{\nu_{f}\nu} X^{*}_{\nu_{i}\nu} X^{*}_{\nu_{f}\nu'} X_{\nu_{i}\nu'}  I(\mu_{K_{i}}, 
\mu_{K_{f}}-\Omega (\nu_{i} -\nu_{f} ),\overline{\epsilon}_{0}-\Omega(\nu_{i}-\nu),\overline{\epsilon}_{0}-\Omega(\nu_{i}-\nu')), \nonumber\\
\end{eqnarray}
while, for processes involving the occupied state, they are given by 
\begin{eqnarray}
\label{cotrate2}
 \gamma^{K_{i}\rightarrow K_{f},1\rightarrow1}_{\nu_{i}\rightarrow\nu_{f}} 
 &=& \frac{1}{2\pi} \Gamma_{K_{i}} \Gamma_{K_{f}} 
\sum_{\nu} \vert X_{\nu_{f}\nu} X_{\nu_{i}\nu} \vert^{2} J(\mu_{K_{i}}, 
\mu_{K_{f}}-\Omega (\nu_{i} -\nu_{f} ),\overline{\epsilon}_{0}+\Omega(\nu_{f}-\nu))\nonumber\\
 &&\hspace{-2.5cm} + \frac{1}{2\pi} \Gamma_{K_{i}} \Gamma_{K_{f}} 
\sum_{\nu'\neq\nu} X_{\nu_{f}\nu} X^{*}_{\nu_{i}\nu} X^{*}_{\nu_{f}\nu'} X_{\nu_{i}\nu'}  I(\mu_{K_{i}}, 
\mu_{K_{f}}-\Omega (\nu_{i} -\nu_{f} ),\overline{\epsilon}_{0}+\Omega(\nu_{f}-\nu),\overline{\epsilon}_{0}+\Omega(\nu_{f}-\nu')), \nonumber\\
\end{eqnarray}
with 
\begin{eqnarray}
 I(E_{1},E_{2},\epsilon_{1},\epsilon_{2}) &=& \frac{1}{\epsilon_{1}-\epsilon_{2}} \frac{1}{\text{e}^{\frac{E_{2}-E_{1}}{k_{\text{B}}T}}-1} 
 \text{Re}\left[ \psi\left(\frac{1}{2}+i\frac{E_{2}-\epsilon_{1}}{2\pi k_{\text{B}}T} \right) - 
                 \psi\left(\frac{1}{2}-i\frac{E_{2}-\epsilon_{2}}{2\pi k_{\text{B}}T}  \right)  \right. \nonumber\\
 && \hspace{4cm} -  \left.    \psi\left(\frac{1}{2}+i\frac{E_{1}-\epsilon_{1}}{2\pi k_{\text{B}}T}  \right) + 
                 \psi\left(\frac{1}{2}-i\frac{E_{1}-\epsilon_{2}}{2\pi k_{\text{B}}T}  \right)  \right] \nonumber\\  \\             
 J(E_{1},E_{2},\epsilon) &=& \frac{1}{2\pi k_{\text{B}}T} \frac{1}{\text{e}^{\frac{E_{2}-E_{1}}{k_{\text{B}}T}}-1} 
 \text{Im}\left[ \psi'\left(\frac{1}{2}+i\frac{E_{2}-\epsilon}{2\pi k_{\text{B}}T}  \right) - 
                 \psi'\left(\frac{1}{2}+i\frac{E_{1}-\epsilon}{2\pi k_{\text{B}}T} \right)  \right] \nonumber \\
\end{eqnarray}
and $\psi(x)$ is the digamma function. Here, off-resonant pair creation processes \cite{Volkovich2011b} 
are associated with rates where $K_{i}=K_{f}$. 
Co-tunneling processes are described by rates where $K_{i}\neq K_{f}$. 
It is straightforward to supplement the rate equations (\ref{explicitsecondorderrates}) 
with these rates \cite{Koch2006}: 
\begin{eqnarray}
\label{explicitcotunnelrates}
0 &=& - \sum_{K,\nu'} \Gamma_{K,0\rightarrow1}^{\nu\rightarrow\nu'} \sigma^{\nu}_{0} 
+ \sum_{K,\nu'} \Gamma_{K,1\rightarrow0}^{\nu'\rightarrow\nu} \sigma^{\nu'}_{1} 
- \sum_{\nu'\neq \nu,K,K'} \gamma_{K\rightarrow K',0\rightarrow0}^{\nu\rightarrow\nu'} 
\sigma^{\nu}_{0} + \sum_{\nu'\neq \nu,K,K'} \gamma_{K\rightarrow K',0\rightarrow0}^{\nu'\rightarrow\nu}  \sigma^{\nu}_{0}, \nonumber \\
&& + \gamma_{\text{bath}} \left( 1 - \text{e}^{-\frac{\Omega}{k_\text{B}T}} \right)^{-1} \left( 
\nu \sigma^{\nu}_{0} - (\nu+1) \sigma^{\nu+1}_{0} \right) 
+ \gamma_{\text{bath}} \left( \text{e}^{\frac{\Omega}{k_\text{B}T}} - 1 \right)^{-1} \left( 
(\nu+1) \sigma^{\nu}_{0} - \nu \sigma^{\nu-1}_{0} \right)  , \nonumber \\
0 &=& - \sum_{K,\nu'} \Gamma_{K,1\rightarrow0}^{\nu\rightarrow\nu'} \sigma^{\nu}_{1} 
+ \sum_{K,\nu'} \Gamma_{K,0\rightarrow1}^{\nu'\rightarrow\nu} \sigma^{\nu'}_{0}
- \sum_{\nu'\neq \nu,K,K'} \gamma_{K\rightarrow K',1\rightarrow1}^{\nu\rightarrow\nu'} 
\sigma^{\nu}_{1} + \sum_{\nu'\neq \nu,K,K'} \gamma_{K\rightarrow K',1\rightarrow1}^{\nu'\rightarrow\nu}  \sigma^{\nu}_{1}, \nonumber \\ 
&& + \gamma_{\text{bath}} \left( 1 - \text{e}^{-\frac{\Omega}{k_\text{B}T}} \right)^{-1} \left( 
\nu \sigma^{\nu}_{1} - (\nu+1) \sigma^{\nu+1}_{1} \right) 
+ \gamma_{\text{bath}} \left( \text{e}^{\frac{\Omega}{k_\text{B}T}} - 1 \right)^{-1} \left( 
(\nu+1) \sigma^{\nu}_{1} - \nu \sigma^{\nu-1}_{1} \right)  , \nonumber \\
\end{eqnarray}
Note that this scheme is valid only for $\Gamma_{K} \ll k_{\text{B}}T,\Omega$ and 
treats cotunneling and off-resonant pair creation processes via virtual tunneling processes. 
Thus, the population of the electronic level is not changed, but the level of vibrational excitation (vide infra).

\subsection{Observables of interest}

With the coefficients of the reduced density matrix, 
we can calculate the vibrational distribution function
\begin{eqnarray}
 p_{\nu} &=& \sigma^{\nu}_{0} + \sigma^{\nu}_{1}. 
\end{eqnarray}
The corresponding average vibrational excitation 
is given by
\begin{eqnarray}
\langle a^{\dagger}a \rangle_{H} &=& \langle a^{\dagger} a \rangle_{\overline{H}} + 
\frac{\lambda^{2}}{\Omega^{2}} \langle d^{\dagger} d \rangle_{H}, \\
&=&   \sum_{\nu} \nu p_{\nu}+ 
\frac{\lambda^{2}}{\Omega^{2}} \langle d^{\dagger} d \rangle_{H}, \nonumber
\end{eqnarray}
and the population of the electronic level by 
\begin{eqnarray}
 \langle d^{\dagger} d \rangle_{H} =\langle d^{\dagger} d \rangle_{\overline{H}} =
\sum_{\nu} \sigma^{\nu}_{1},
\end{eqnarray}
where the indices $H$/$\overline{H}$ indicate the Hamiltonian that is used to evaluate 
the respective expectation value. It should be noted at this point that the current, which 
is flowing through the electronic level in the limit $\lambda\rightarrow0$,  
is the same as in the non-interacting case \cite{Hartle2011}.

\section{Results}
\label{results}

In the following, we discuss vibrationally coupled electron transport, focusing on the 
counterintuitive phenomenon that, in the steady state regime, 
a larger level of vibrational excitation can be obtained 
for systems with a weaker electronic-vibrational coupling 
\cite{Mitra04,Semmelhack,Avriller2010,Hartle2011,Ankerhold2011}. 
To this end, we discuss the basics of the phenomenon in Sec.\ \ref{Basics}, using Born-Markov theory. 
We continue to investigate the weak coupling limit, $\lambda\rightarrow0$, 
where the largest level of vibrational excitation is obtained. 
Thereby, we consider the influence of thermal broadening and hybridization with the electrodes in Sec.\ \ref{thermBroad}. 
In Sec.\ \ref{BroadCoTun}, we employ the cotunneling rates (\ref{cotrate}) and (\ref{cotrate2}) 
and study the influence of inelastic cotunneling and off-resonant pair creation processes. 
The effect of broadening due to a heat bath is discussed 
in Sec.\ \ref{heatbathsec}.

\subsection{Basics of vibrationally coupled transport}
\label{Basics}

\begin{figure}
\resizebox{\newwidth}{\newheight}{
\includegraphics{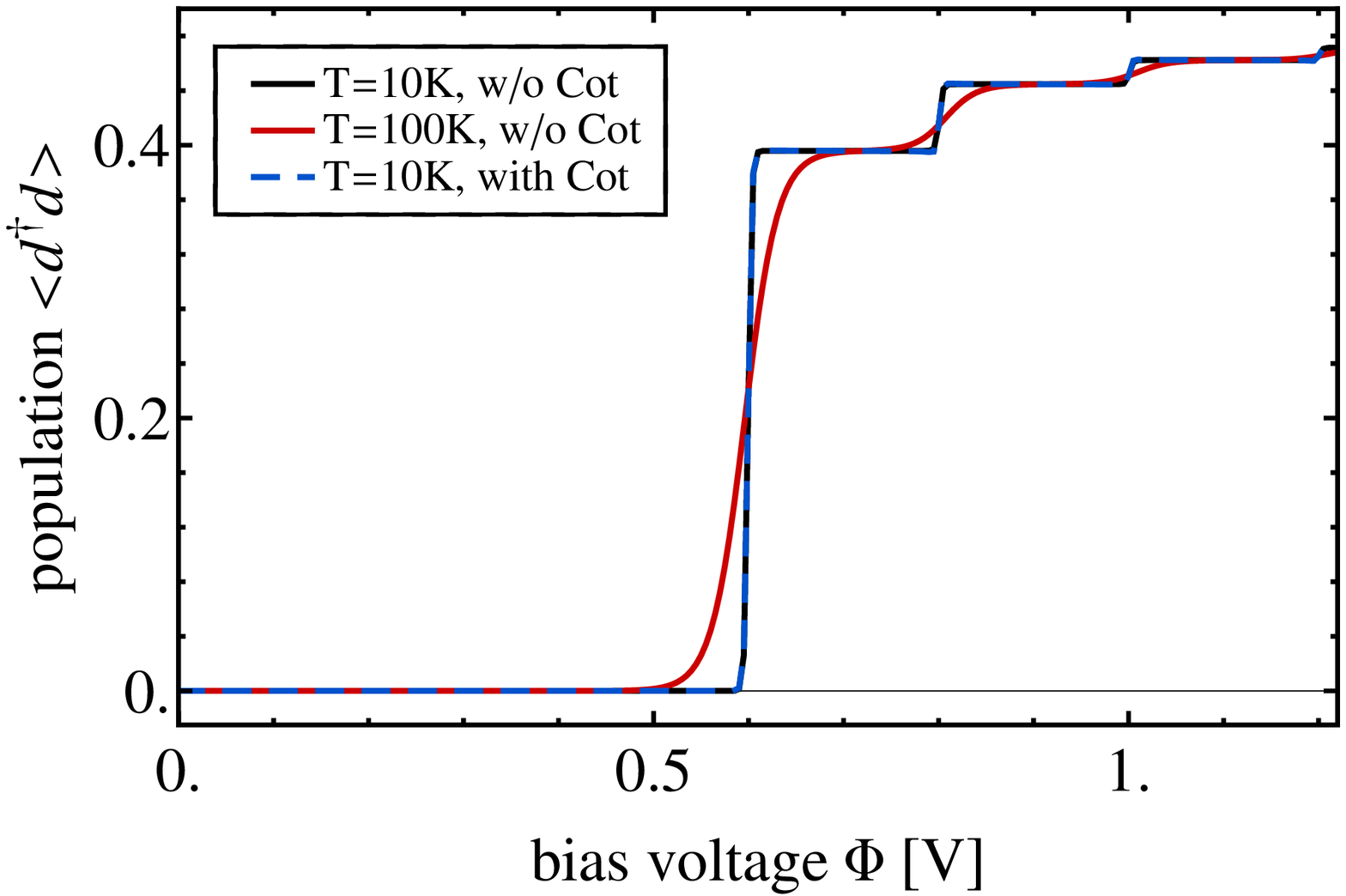}
}
\resizebox{\newwidth}{\newheight}{
\includegraphics{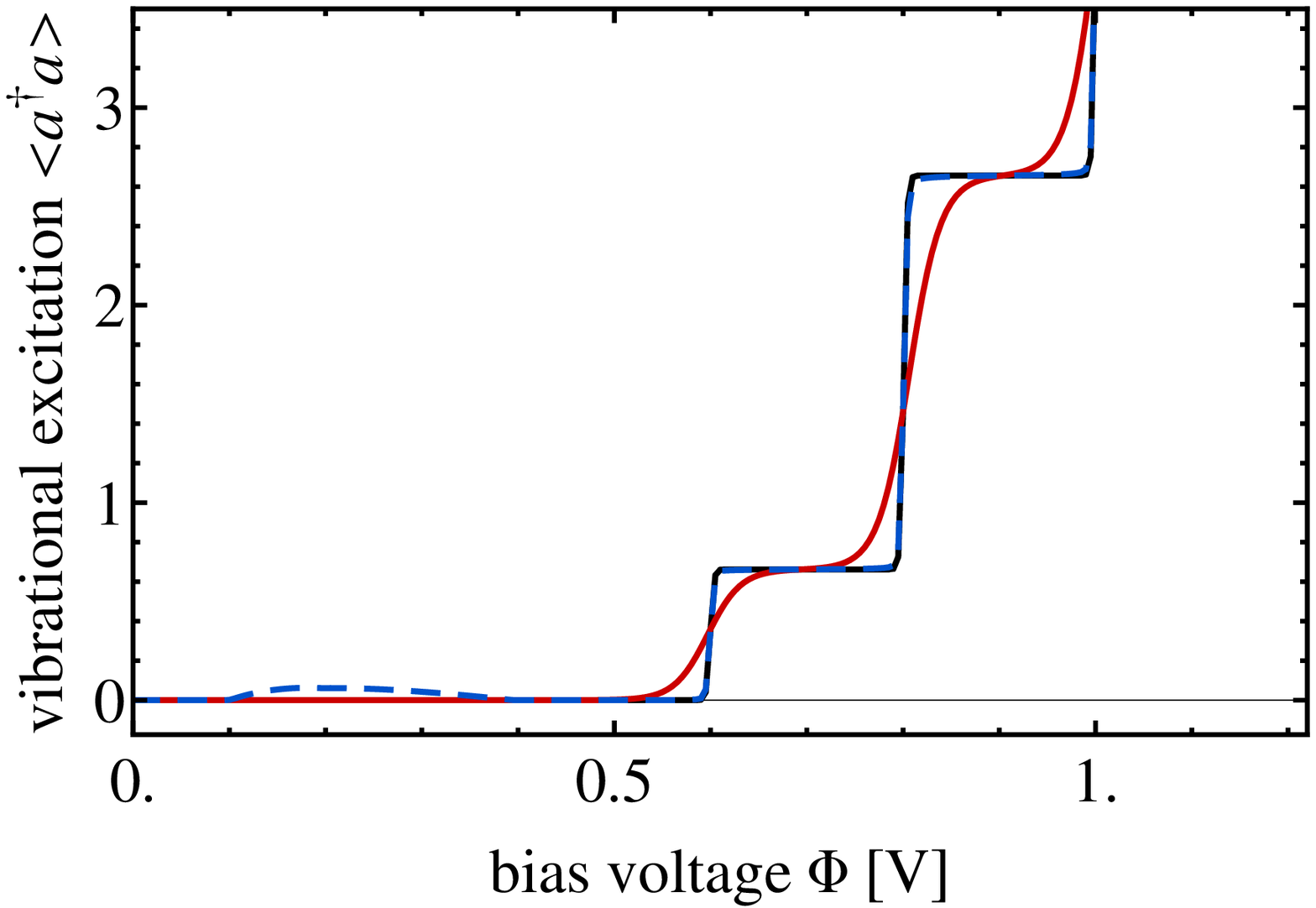}
}
\caption{\label{varbias} Electronic population and vibrational excitation characteristics 
of our model molecular junction (all parameters are given in the text).}
\end{figure}

We employ a minimal model for 
a molecular junction. It comprises a single electronic state that is located 
$\overline{\epsilon}_{0}=0.3$\,eV above the Fermi level of the junction. The coupling of this state 
to the electrodes is characterized by the hybridization strengths $\Gamma=\Gamma_{K}=0.2$\,meV, 
which are smaller than the thermal broadening induced by the temperature in the leads $T=10$\,K. 
The vibrational degrees of freedom are subsumed in a single vibrational mode with frequency $\Omega=100$\,meV. 
It is coupled to the electronic state by an intermediate coupling strength of $\lambda=50$\,meV. 
For the present purpose, we do not consider a coupling of the mode to a heat bath, \emph{i.e.}\ $\gamma_{\text{bath}}=0$. 
The electronic population and vibrational excitation characteristics of this model molecular junction 
are depicted by the solid black lines in Fig.\ \ref{varbias} as functions of the applied bias voltage. 
They have been obtained using the rate equations (\ref{explicitsecondorderrates}) 
and can be understood as follows.

\begin{figure}
\begin{center}
\begin{tabular}{llll}
(a)&(b)&(c)&(d)\\
\resizebox{\newwidthprime}{\newheightprime}{
\includegraphics{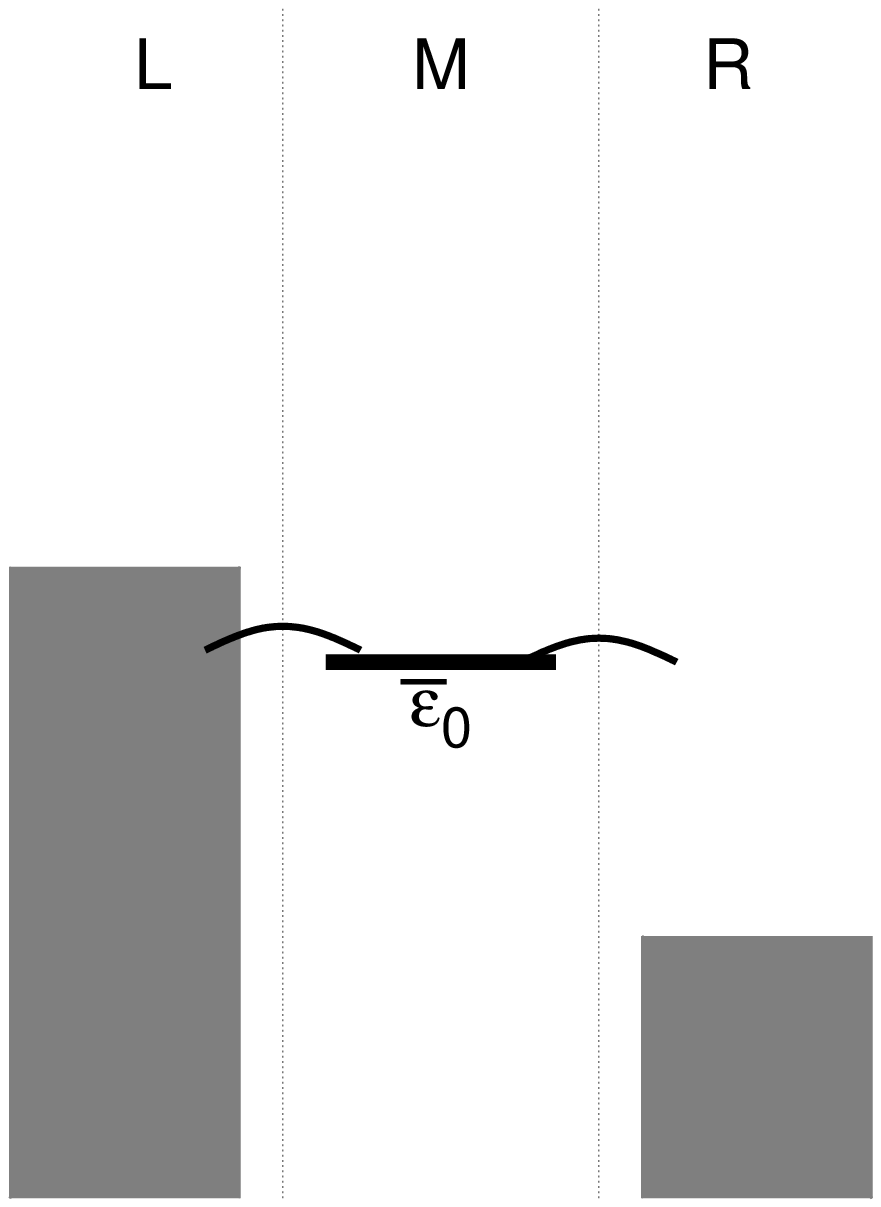}
}&
\resizebox{\newwidthprime}{\newheightprime}{
\includegraphics{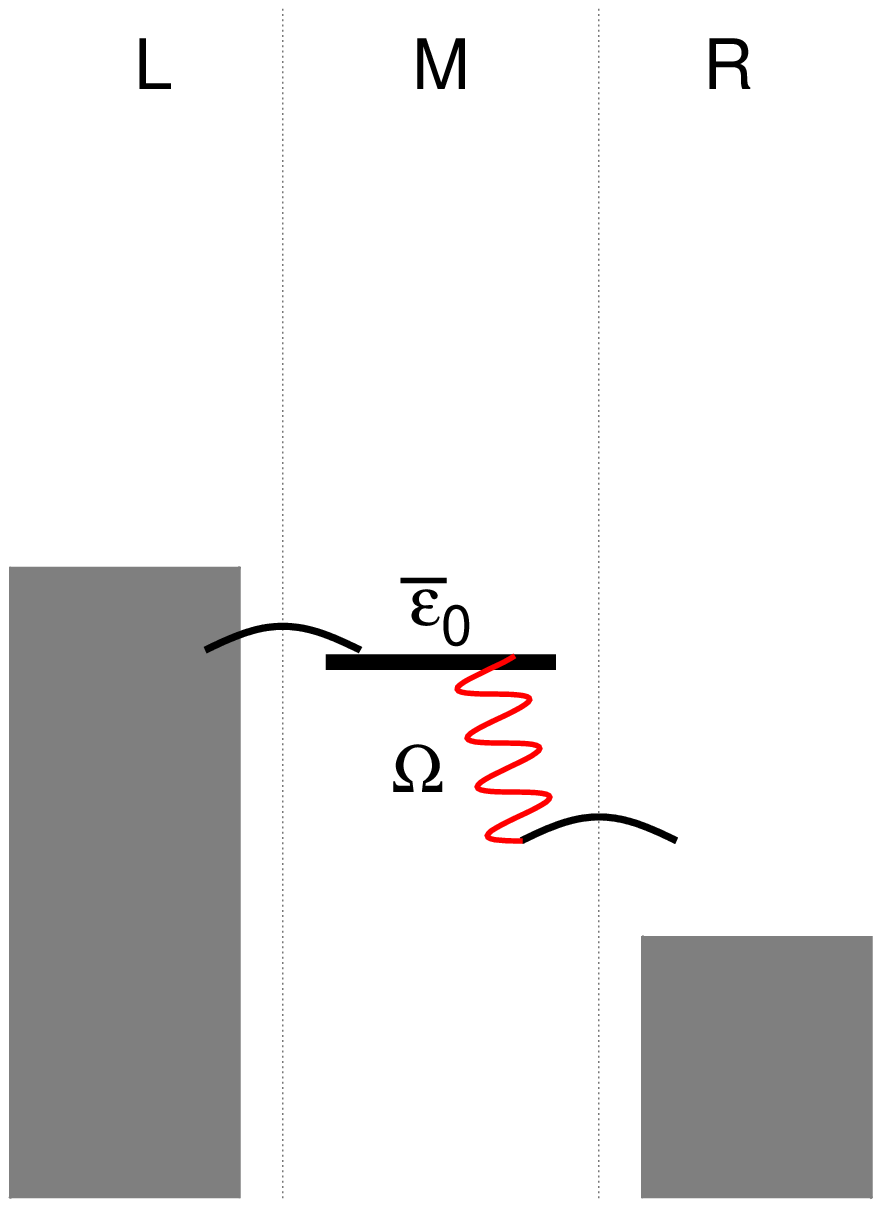}
}&
\resizebox{\newwidthprime}{\newheightprime}{
\includegraphics{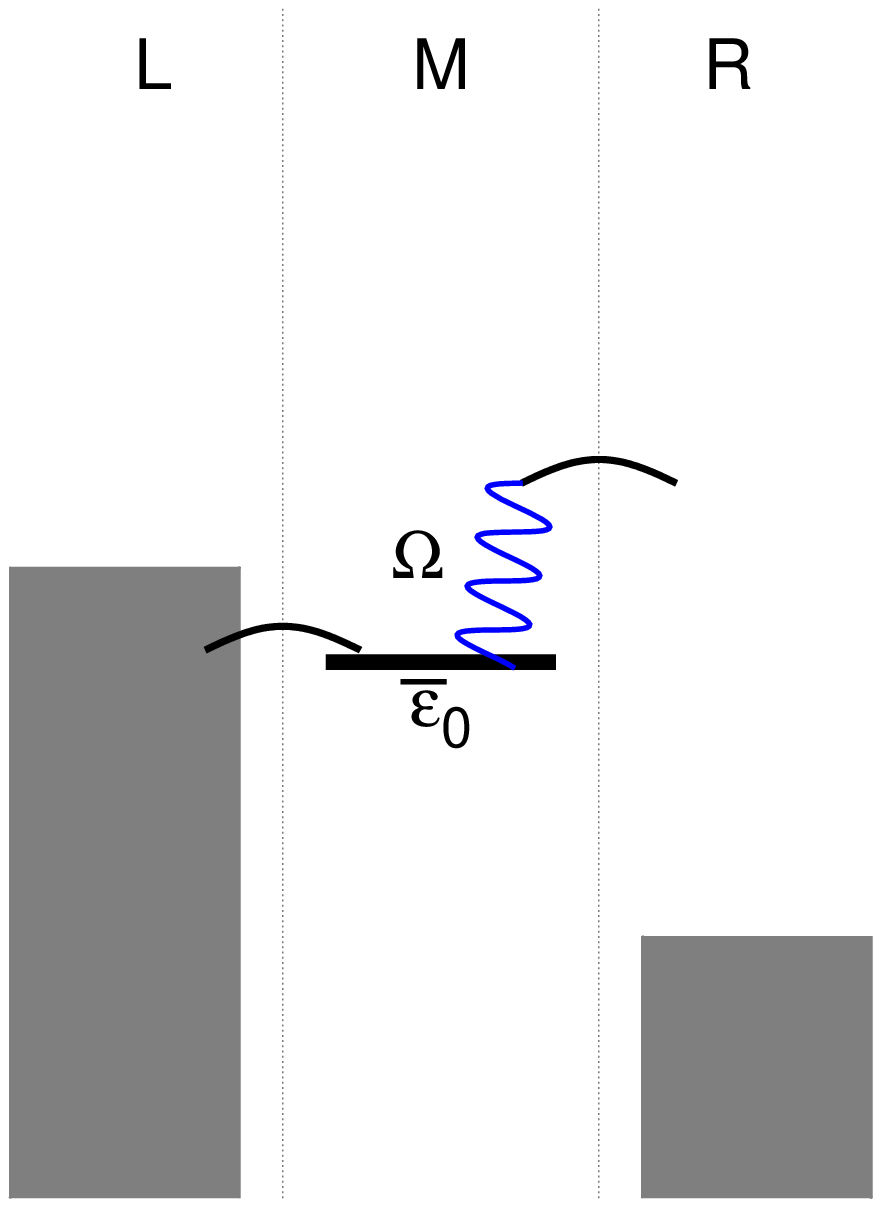}
}&
\resizebox{\newwidthprime}{\newheightprime}{
\includegraphics{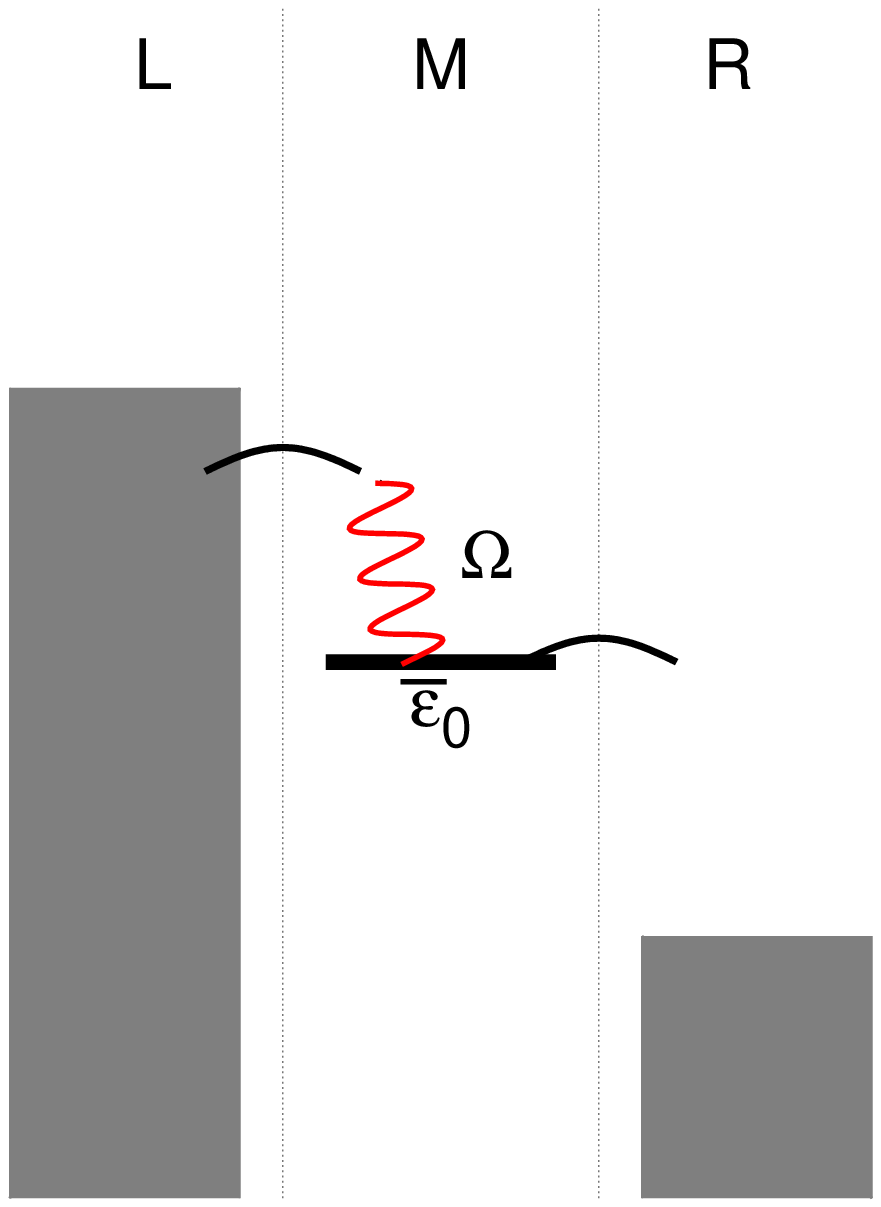}
}
\\ 
\end{tabular}
\end{center}
\caption{\label{basmech} Graphical representation of example processes for 
sequential tunneling. 
Panel (a) depicts a sequential tunneling process where an electron 
from the left lead tunnels to the right lead in two resonant tunneling processes. 
In panel (b)/(c) the latter of the processes is accompanied by an 
excitation/deexcitation process where the vibrational mode is excited/deexcited 
by a quantum of vibrational energy. 
The processes (a)--(c) become active at the same bias voltage, 
that is for $e\Phi\approx2\overline{\epsilon}_{0}$. 
The resonant excitation process shown in Panel (d) requires  
a higher bias voltage, that is  
$e\Phi\gtrsim2(\overline{\epsilon}_{0}+\Omega)$.}
\end{figure}

For bias voltages $e\Phi<2\overline{\epsilon}_{0}$, the polaron-shifted 
electronic level is far above the chemical potentials in the electrodes and, therefore, exhibits no 
significant population. At $e\Phi\approx2\overline{\epsilon}_{0}$, it becomes populated 
because the chemical potential in the left lead is high enough to allow for 
resonant tunneling events (which are graphically depicted Figs.\ \ref{basmech}a--c). 
Its population increases even further at $e\Phi\approx2(\overline{\epsilon}_{0}+m\Omega)$ with $m\in\mathbb{N}$  
due to the onset of inelastic tunneling processes, which involve the excitation 
of the vibrational degree of freedom by $m$ vibrational quanta during the tunneling process 
from the left electrode onto the molecule (see Fig.\ \ref{basmech}d for a process with 
a single vibrational quantum). 
In general, the population of the electronic level is given by 
the ratio of the time scales to populate and depopulate it. 
For $\Gamma_{\text{L}}=\Gamma_{\text{R}}$, these time scales are 
determined by the number and probability of tunneling processes with 
respect to the left and the right lead, respectively. For $e\Phi>2\overline{\epsilon}_{0}$, 
this ratio increases whenever a tunneling 
process from the left lead onto the electronic level becomes available at 
$e\Phi\approx2(\overline{\epsilon}_{0}+m\Omega)$. As the probability associated with these processes 
decreases, the heights of the corresponding steps in the population characteristics 
become smaller.

\begin{figure}
\begin{center}
\begin{tabular}{lll}
(a)&(b)&(c)\\
\resizebox{\newwidthprime}{\newheightprime}{
\includegraphics{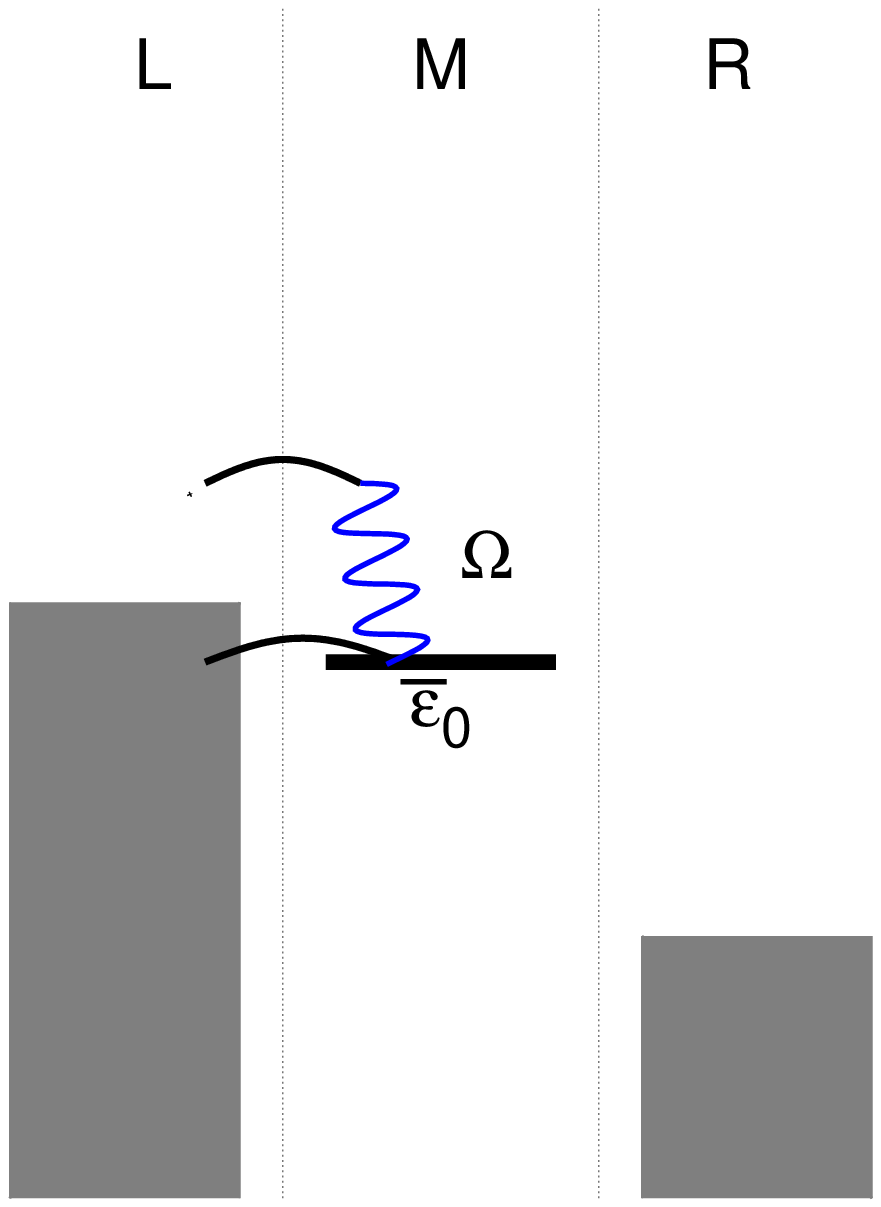}
}
&
\resizebox{\newwidthprime}{\newheightprime}{
\includegraphics{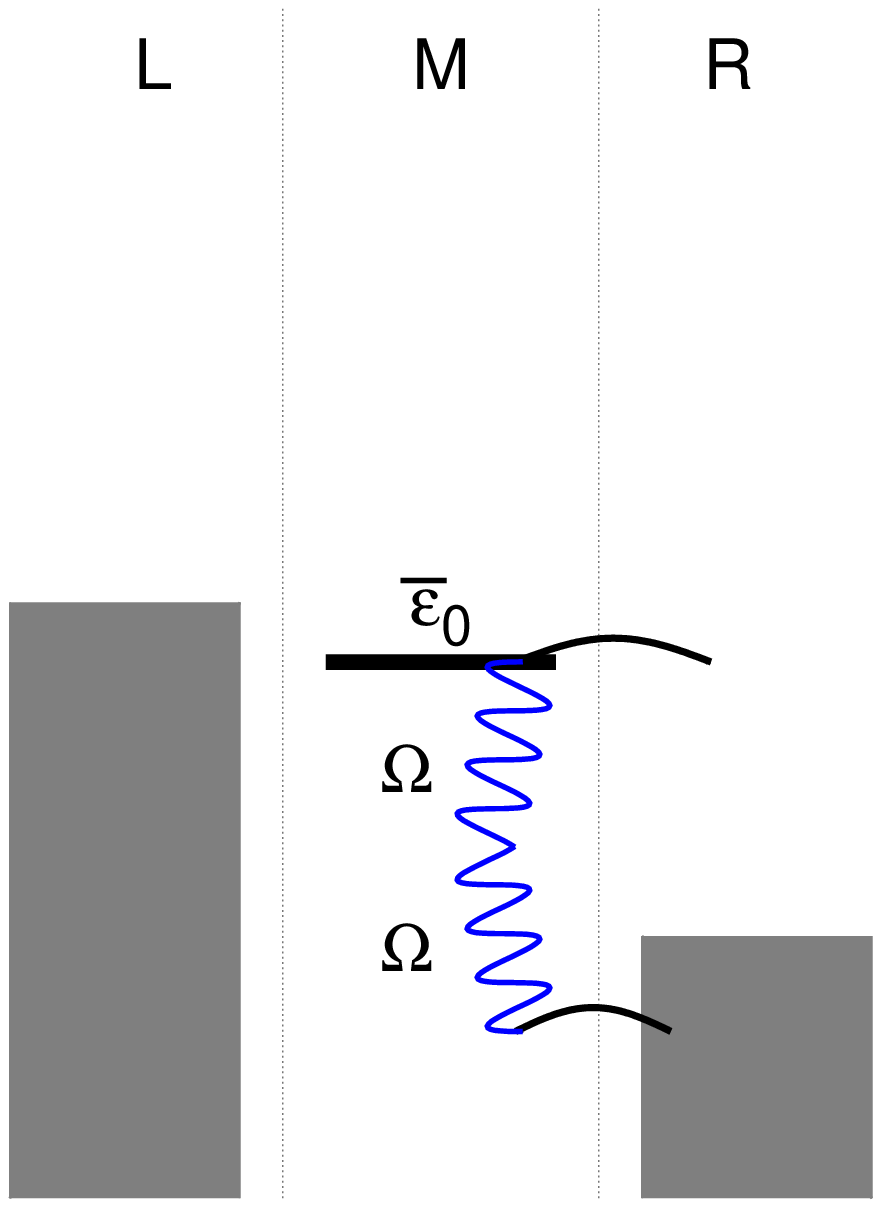}
}
&
\resizebox{\newwidthprime}{\newheightprime}{
\includegraphics{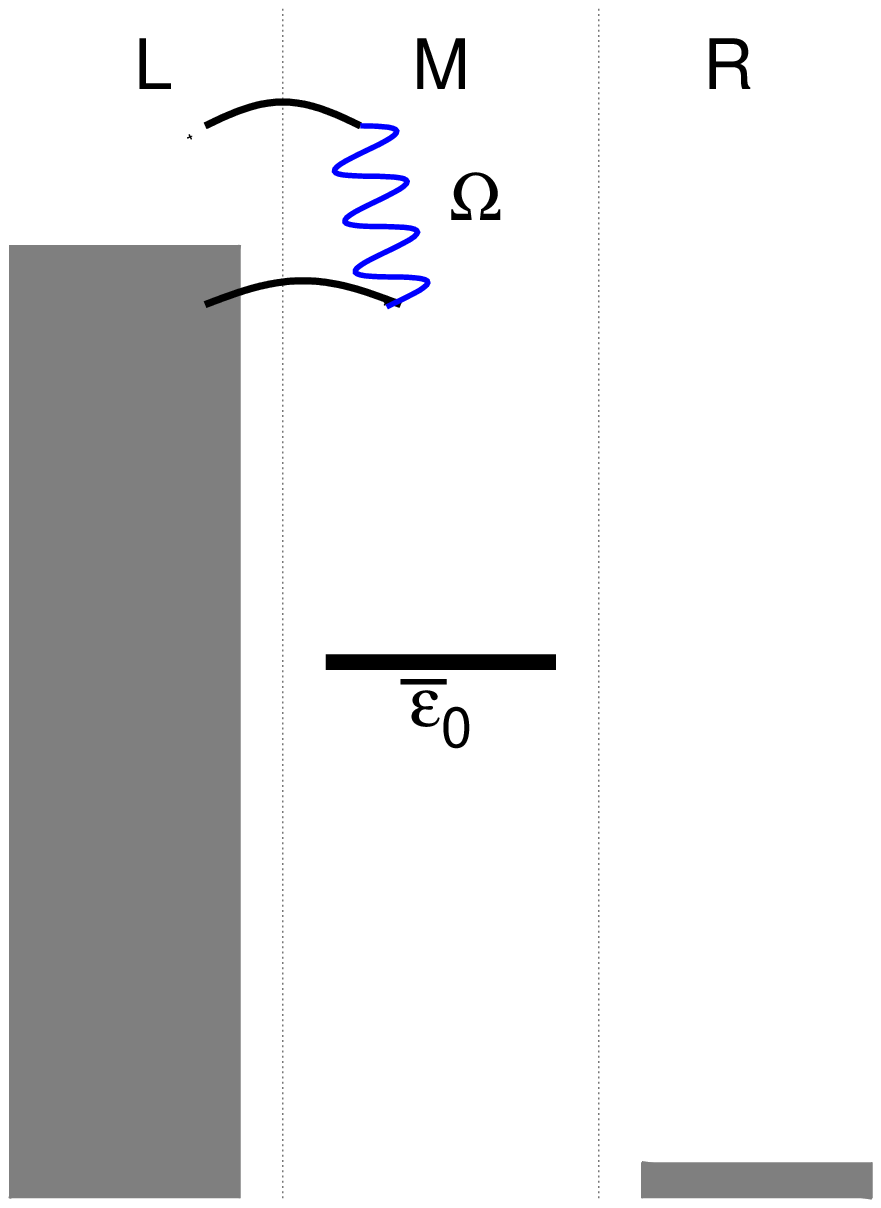}
}
\\ 
\end{tabular}
\end{center}
\caption{\label{el-h-pair-creation} Graphical representation of example 
electron-hole pair creation processes. 
Panel (a) shows a resonant pair creation process, 
where an electron tunnels from the left lead onto 
the molecular bridge and back to the left lead in two sequential tunneling events, 
absorbing a quantum of vibrational energy. 
Panel (b) represents a resonant pair creation process with respect to the right lead, 
involving two vibrational quanta, which, for weak electronic-vibrational coupling, 
is less favorable than processes with only one vibrational quantum. 
Panel (c) depicts an off-resonant pair creation processes. 
These processes represent the pair creation analog of deexcitation 
via inelastic co-tunneling (cf.\ Fig.\ \ref{IETSprocesses}c). 
}
\end{figure}

Similar to the electronic population, the level of vibrational excitation increases 
at the same bias voltages. This quantity, however, is determined by the ratio between vibrational 
excitation and deexcitation processes. This includes both transport 
(cf.\ Fig.\ \ref{basmech}a--d) and electron-hole pair creation processes 
(cf.\ Fig.\ \ref{el-h-pair-creation}a--b). While the former are 
sufficient to understand the electronic population on a qualitative level, 
the vibrational excitation characteristics can only be understood if 
the effect of electron-hole pair creation processes is taken into account 
\cite{Hartle2010b,Hartle2011,Hartle2013}.

In contrast to the electronic population, the level of vibrational excitation 
increases indefinitely as $\Phi\rightarrow\infty$ \cite{Hartle2011}. In this limit, 
electron-hole pair creation processes are completely suppressed 
and only transport processes are available. 
Thereby, each transport-induced excitation process is partnered by 
a corresponding deexcitation process (the partner of the process shown in Fig.\ \ref{basmech}b 
is the one depicted by Fig.\ \ref{basmech}c). Thus, the ratio between excitation and 
deexcitation processes is equal. This triggers a random walk through 
the ladder of the vibrational states \cite{Semmelhack} such that all vibrationally excited states 
become equally populated. As a result, the 
level of vibrational excitation increases indefinitely \cite{Semmelhack,Hartle2011}.

At finite bias voltages, electron-hole pair creation processes are active. 
However, they become successively suppressed if $e\Phi$ exceeds the values 
$2(\overline{\epsilon}_{0}+m\Omega)$ where $m\in\mathbb{Z}$ (\emph{e.g.},  
the process shown in Fig.\ \ref{el-h-pair-creation}a is suppressed for $e\Phi>2(\overline{\epsilon}_{0}+\Omega)$).  
For small coupling strengths $\lambda$, 
these processes are more important the less vibrational quanta $m$ they involve. 
Accordingly, deexcitation of the vibrational mode via pair creation processes is most pronounced in the range of 
bias voltages from $e\Phi=2(\overline{\epsilon}_{0}-\Omega)$ to $e\Phi=2(\overline{\epsilon}_{0}+\Omega)$. 
Outside this voltage range, resonant pair creation processes require at least two vibrational quanta. 
At lower bias voltages, $e\Phi<2\overline{\epsilon}_{0}$, resonant deexcitation processes are more favorable 
than resonant excitation processes. Thus, only at higher bias voltages, 
$e\Phi>2(\overline{\epsilon}_{0}+\Omega)$, a significant level of vibrational excitation can develop. 
Thereby, for increasing bias voltages, the ratio between excitation and deexcitation processes becomes equal 
due to the successive suppression of pair creation processes. 
As a result, the step heights in the vibrational excitation characteristics do not tend to zero 
such as, for example, the steps in the corresponding population characteristic of the electronic level 
but 
increase with the applied bias voltage \cite{Hartle2010b,Hartle2011}.

\begin{figure}
\resizebox{\newwidth}{\newheight}{
\includegraphics{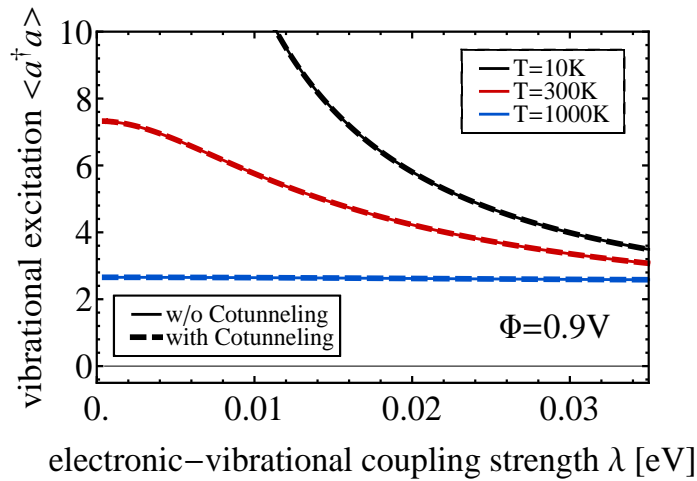}
}
\caption{\label{vartemp} Level of vibrational excitation at $e\Phi=2(\overline{\epsilon}_{0}+3\Omega/2)$ 
as a function of the electronic-vibrational coupling strength $\lambda$. We have obtained very similar results 
throughout the range of bias 
voltages $2(\overline{\epsilon}_{0}+\Omega+k_{\text{B}}T)<e\Phi<2(\overline{\epsilon}_{0}+2\Omega-k_{\text{B}}T)$ 
(data not shown).}
\end{figure}

As the pair creation processes that become suppressed at 
$e\Phi\approx2(\overline{\epsilon}_{0}+m\Omega)$ with $m\in\mathbb{N}$ are more important 
for smaller coupling strengths $\lambda$, the level of vibrational excitation is higher 
when the coupling between the electronic state and the vibrational degree of freedom is weaker. 
This counterintuitive phenomenon has been reported by a number of groups 
\cite{Mitra04,Semmelhack,Avriller2010,Hartle2011,Ankerhold2011}. 
It is demonstrated in Fig.\ \ref{vartemp} by the solid black line, which shows the level of 
vibrational excitation of our model molecular junction as a function of the electronic-vibrational coupling strength 
at a fixed bias voltage $e\Phi=2(\overline{\epsilon}_{0}+3\Omega/2)=0.9$\,V. It can be seen that 
the level of vibrational excitation increases as $\lambda\rightarrow0$. 
This is the situation discussed in Ref.\ \cite{Hartle2011}. There, the infinite bias limit was 
shown to be equal to the limit of a vanishing electronic-vibrational coupling in the sense that deexcitation 
due to electron-hole pair creation processes becomes suppressed. This finding does not include 
the effect of broadening. 
In the following sections, we extend these considerations to account for thermal broadening 
and the hybridization with the electrodes (Sec.\ \ref{thermBroad}), 
the effect of higher-order processes (Sec.\ \ref{BroadCoTun}) and the coupling of the vibrational mode 
to a thermal heat bath (Sec.\ \ref{heatbathsec}).

\subsection{Broadening due to temperature and hybridization with the electrodes}
\label{thermBroad}

The effect of thermal broadening can be demonstrated by recalculating the 
transport characteristics of our model molecular junction with an increased 
temperature: $T=100$\,K. The corresponding electronic population and vibrational 
excitation characteristics are depicted by the solid red lines in Fig.\ \ref{varbias}. 
As compared to the black lines, the steps are much broader. Otherwise, no qualitative changes 
appear. This is different considering the limit $\lambda\rightarrow0$. 
The red line in Fig.\ \ref{vartemp} shows a clear tendency towards a finite 
level of vibrational excitation for vanishing electronic-vibrational coupling (for $T=10$\,K, 
this level is just much higher). 
The physical origin of this behavior can be explained in terms of electron-hole pair creation 
processes. Due to thermal broadening or thermal fluctuations, the suppression of electron-hole pair creation 
processes at $e\Phi>2(\overline{\epsilon}_{0}+m\Omega)$ is not complete. The corresponding deexcitation 
mechanism is not very efficient, but sufficient to keep the vibrational excitation of the junction 
at a finite level.

This behavior can be analyzed more rigorously using analytic arguments. To this end, 
we employ the Born-Markov master equation (\ref{explicitsecondorderrates}). 
In the limit $\lambda\rightarrow0$ and bias voltages $e\Phi\gtrsim2(\overline{\epsilon}_{0}+3\Omega/2)$, 
the resulting rate equations (\ref{explicitsecondorderrates}) can be simplified to 
\begin{eqnarray}
\label{thermMEqu}
0 &=& \Gamma_{R} \sigma^{\nu}_{1} - \Gamma_{\text{L}}  \sigma^{\nu}_{0} 
+ \Gamma_{R}  \lambda^{2} \left( (\nu+1) 
\sigma^{\nu+1}_{1} - (2\nu+1) \sigma^{\nu}_{1} + \nu \sigma^{\nu-1}_{1} \right) \nonumber\\
 && + (1-f_{\text{L}}(\epsilon_{0}+\Omega)) 
\Gamma_{\text{L}} \lambda^{2} (\nu+1)
\sigma^{\nu}_{0} \nonumber\\
&& + (1-f_{\text{L}}(\epsilon_{0}+\Omega))   
\Gamma_{\text{L}} \lambda^{2} (\nu+1) 
\sigma^{\nu+1}_{1}  
\end{eqnarray}
and 
\begin{eqnarray}
\label{thermMEqu2}
0 &=& \Gamma_{\text{L}} \sigma^{\nu}_{0} - \Gamma_{R}  \sigma^{\nu}_{1} 
+  \Gamma_{\text{L}} \lambda^{2} \left( (\nu+1)  \sigma^{\nu+1}_{0} -  
(2\nu+1)\sigma^{\nu}_{0} + \nu  \sigma^{\nu-1}_{0} \right)   \nonumber\\
&& - (1-f_{\text{L}}(\epsilon_{0}+\Omega)) 
\Gamma_{\text{L}} \lambda^{2} \nu  \sigma^{\nu}_{1}  \nonumber\\
&& -  (1-f_{\text{L}}(\epsilon_{0}+\Omega)) \Gamma_{\text{L}}
\lambda^{2} \nu  \sigma^{\nu-1}_{0} 
\end{eqnarray}
where we used the approximations $X_{\nu\nu}\approx1-\lambda^{2}(2\nu+1)/2$, 
$X_{\nu\nu+1}\approx\lambda\sqrt{\nu+1}$ and evaluated each Fermi function by either $0$ or $1$ 
except for $f_{\text{L}}(\epsilon_{0}+\Omega)$, which allows us to take into account 
the most pronounced effect of thermal broadening. 
The corresponding term describes deexcitation via the electron-hole pair creation process 
depicted in Fig.\ \ref{el-h-pair-creation}a. 
For $T=0$, the equations (\ref{thermMEqu}) and (\ref{thermMEqu2}) 
are the same as Eqs.\ (22) and (23) in Ref.\ \onlinecite{Hartle2011}. 
Note that contributions from the $\lambda$-dependence of $\overline{\epsilon}_{0}$ can be neglected, 
because the corresponding terms are of higher order or because the derivatives of the corresponding 
Fermi functions are evaluated at energies far from the chemical potentials.

From equations (\ref{thermMEqu}) and (\ref{thermMEqu2}), we can infer that 
$\Gamma_{\text{R}}\sigma_{1}^{\nu}-\Gamma_{\text{L}}\sigma_{0}^{\nu}=\mathcal{O}(\lambda^{2})$. 
This allows us to further simplify the above equations by replacing $\Gamma_{\text{L}}\sigma_{0}^{\nu}$ 
with $\Gamma_{\text{R}}\sigma_{1}^{\nu}$ (if these terms are accompanied by a factor $\lambda^{2}$). 
The sum of the resulting equations can be written as 
\begin{eqnarray}
0 &=& + 2 \Gamma_{\text{R}}   \left( (\nu+1) 
\sigma^{\nu+1}_{1} - (2\nu+1) \sigma^{\nu}_{1} + \nu \sigma^{\nu-1}_{1} \right) \\
 && + (1-f_{\text{L}}(\epsilon_{0}+\Omega)) \left( 
 \Gamma_{\text{R}}  (\nu+1) \sigma^{\nu}_{1} + \Gamma_{\text{L}}  (\nu+1) 
\sigma^{\nu+1}_{1} - \Gamma_{\text{L}}  \nu  \sigma^{\nu}_{1} 
-  \Gamma_{\text{R}} \nu  \sigma^{\nu-1}_{1} \right). \nonumber
\end{eqnarray}
The solution of this recurrence relation can be obtained with the 
ansatz $\sigma_{1}^{\nu+1}=\alpha\sigma_{1}^{\nu}$. 
It is given by 
\begin{eqnarray}
 \sigma_{1}^{\nu+1} &=& \frac{ 1+f_{\text{L}}(\epsilon_{0}+\Omega) }{2+ 
 \frac{\Gamma_{\text{L}}}{\Gamma_{\text{R}}}
 (1-f_{\text{L}}(\epsilon_{0}+\Omega)) } \sigma_{1}^{\nu}.  
\end{eqnarray}
Similarly, one can obtain 
\begin{eqnarray}
 \sigma_{0}^{\nu+1} &=& \frac{ 1+f_{\text{L}}(\epsilon_{0}+\Omega) }{2+
 \frac{\Gamma_{\text{R}}}{\Gamma_{\text{L}}}
 (1-f_{\text{L}}(\epsilon_{0}+\Omega)) } \sigma_{0}^{\nu}. 
\end{eqnarray}
For $\Gamma_{\text{R}}=\Gamma_{\text{L}}$, 
the corresponding level of vibrational excitation is given by 
\begin{eqnarray}
 \langle a^{\dagger}a \rangle &=& (1-f_{\text{L}}(\epsilon_{0}+\Omega))^{-1} - \frac{1}{2},  
\end{eqnarray}
which agrees very well with the numerical values that are shown in Fig.\ \ref{vartemp} 
in the limit $\lambda\rightarrow0$.

This analysis demonstrates the importance of deexcitation via electron-hole pair creation processes. 
It represents an extension of the discussion given in Ref.\ \cite{Hartle2011}, where 
thermal broadening was not taken into account. In that case, 
it was shown that the limit $\lambda\rightarrow0$ 
is equivalent to the high bias limit $\Phi\rightarrow\infty$, that is 
the level of vibrational excitation increases indefinitely in the limit 
$\lambda\rightarrow0$. 
Including thermal broadening, however, 
the level of vibrational excitation does not increase indefinitely but 
tends to a finite value, which is determined by the probability that resonant 
pair creation processes can occur. 
This finding is similar to the one discussed in Ref.\ \cite{Hartle2011} for lower voltages,  
$2(\epsilon_{0}+\Omega)>e\Phi>2(\epsilon_{0})$.

So far, we have discussed thermal broadening. 
The electronic level exhibits, in addition, broadening due to the coupling to the 
electrodes. The corresponding width is given by the sum $\Gamma_{\text{L}}+\Gamma_{\text{R}}$. 
This type of broadening facilitates the same deexcitation mechanisms by resonant 
electron-hole pair creation processes as thermal broadening 
and can be expected to have the same influence on the level of 
vibrational excitation in the weak coupling limit. 
We therefore conclude that the level of vibrational excitation 
stays finite in the limit $\lambda\rightarrow0$. We would like emphasize that it 
is, nevertheless, non-zero. 
The latter statement is important, as it demonstrates that the level of vibrational excitation 
can be non-analytic in nonequilibrium situations.

\subsection{Broadening due to cotunneling}
\label{BroadCoTun}

\begin{figure}
\begin{center}
\begin{tabular}{lll}
(a) & (b) & (c)\\
\resizebox{\newwidthprime}{\newheightprime}{
\includegraphics{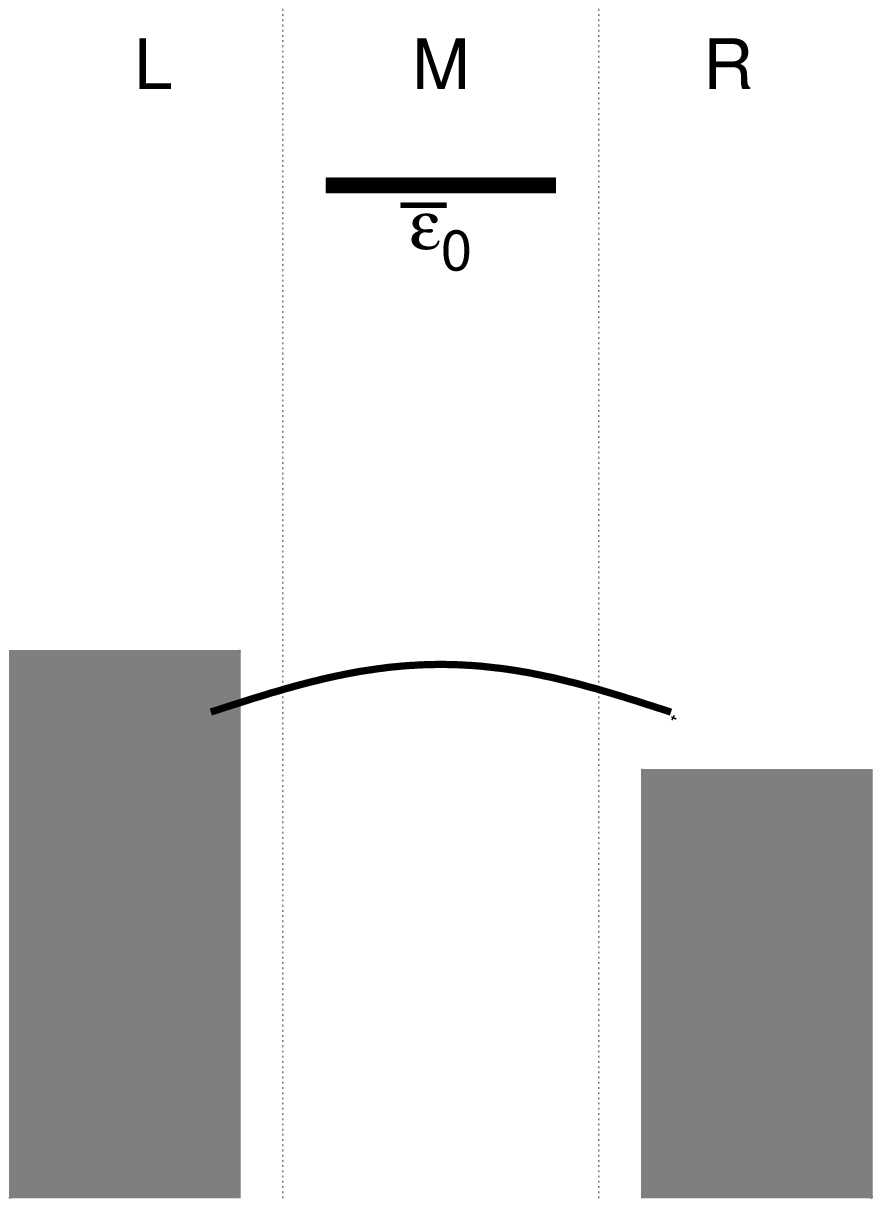}
}&
\resizebox{\newwidthprime}{\newheightprime}{
\includegraphics{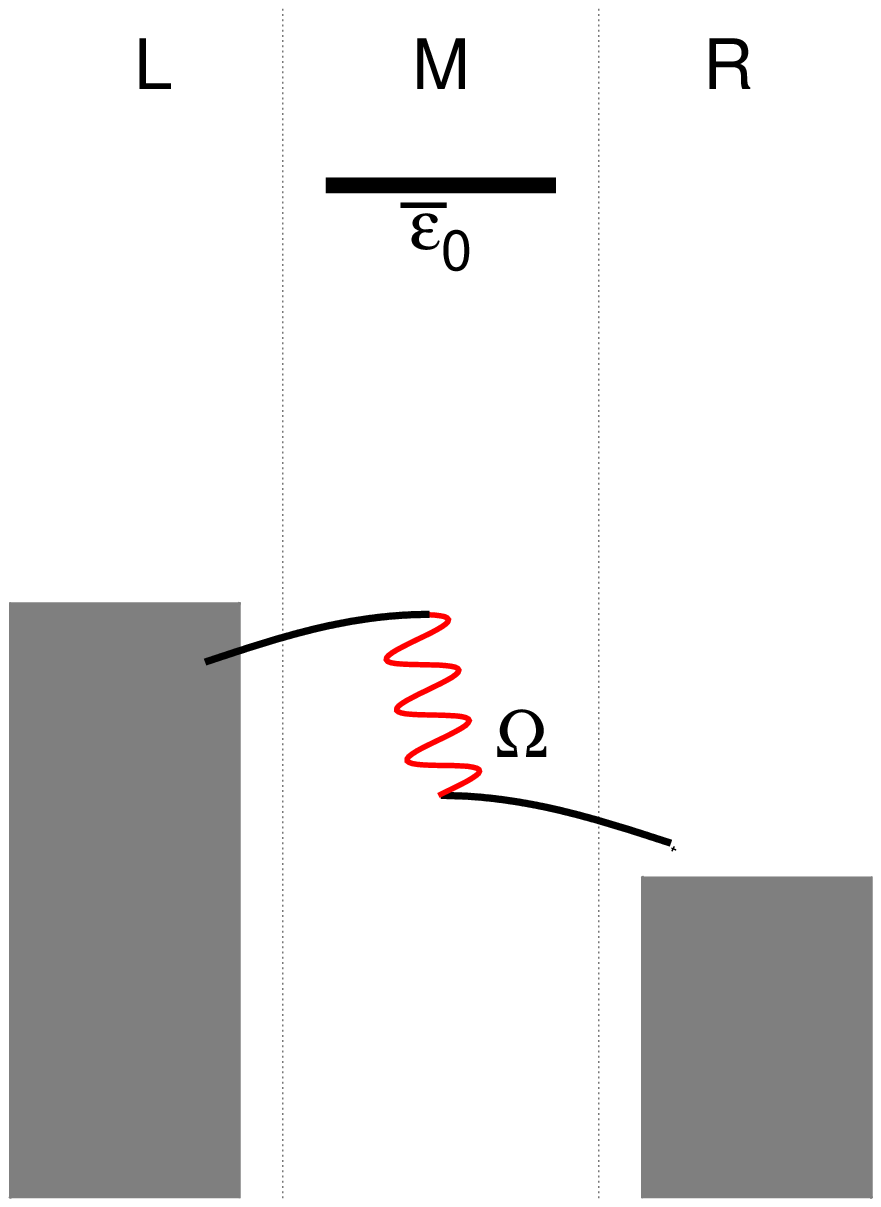}
}&
\resizebox{\newwidthprime}{\newheightprime}{
\includegraphics{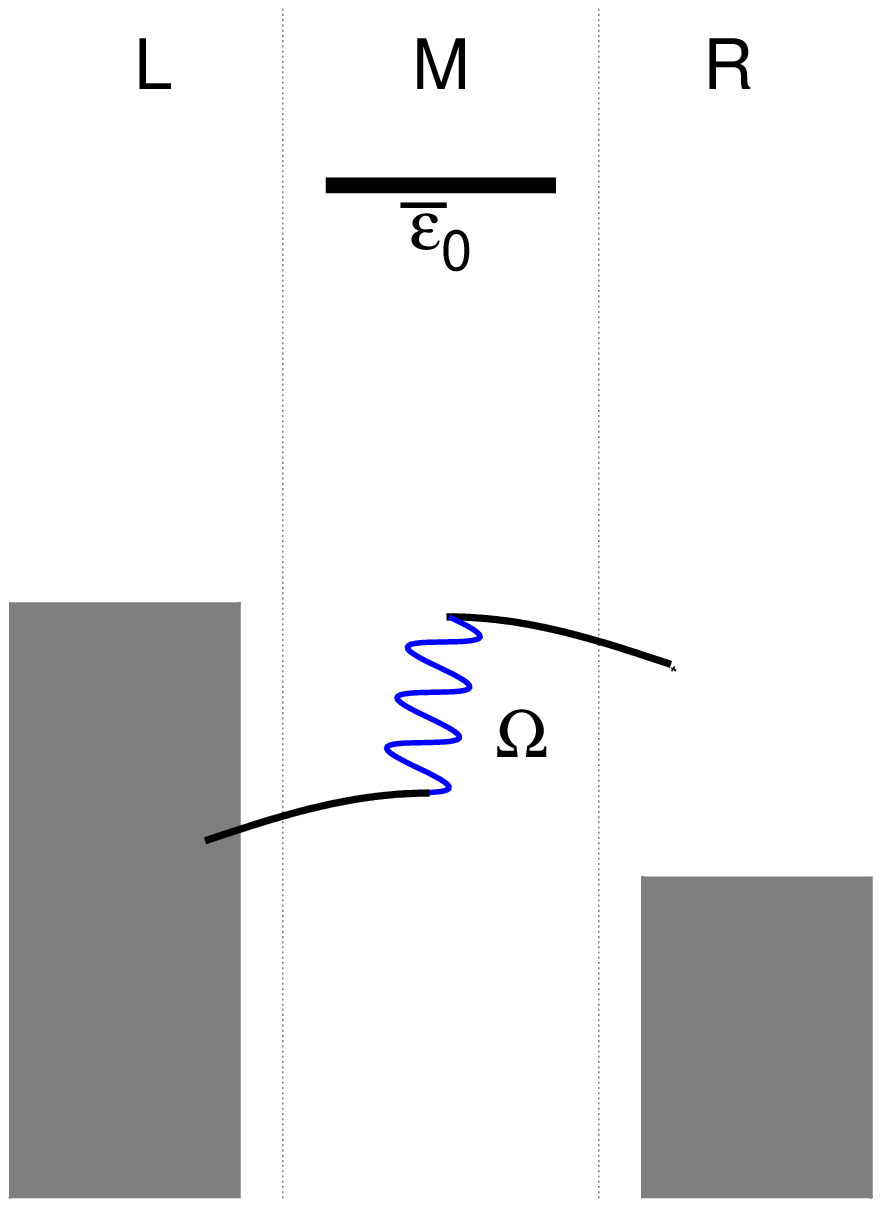}
}\\ 
\end{tabular}
\end{center}
\caption{\label{IETSprocesses} Graphical representation of 
co-tunneling processes. Panel (a) shows direct tunneling of an electron from the 
left to the right lead. Panel (b) and (c) depict inelastic co-tunneling, where the 
vibrational degree of freedom is excited and deexcited by a single quantum of 
vibrational energy, respectively. }
\end{figure}

As we have seen, broadening facilitates resonant electron-hole pair creation 
processes and that these processes are particularly important for deexcitation of the vibrational mode 
in the limit $\lambda\rightarrow0$. 
We now investigate the role of off-resonant pair creation processes (an example 
is depicted in Fig.\ \ref{el-h-pair-creation}c) and inelastic co-tunneling processes 
(see Fig.\ \ref{IETSprocesses}), neglecting, at the same time, broadening effects. 
Recently, a direct comparison of Born-Markov and nonequilibrium Green's function 
results \cite{Volkovich2011b} pointed to a deexcitation mechanism 
via off-resonant pair creation processes which is important 
whenever resonant processes become suppressed. This is the case, for example, 
at high bias voltages where these processes result in a reduced level of vibrational excitation. 
Therefore, it is interesting to check the importance of these processes 
in the limit $\lambda\rightarrow 0$.

We demonstrate the effect of inelastic co-tunneling and off-resonant pair creation processes 
by the dashed blue lines in Fig.\ \ref{varbias}. It shows the 
electronic population and the vibrational excitation characteristics 
of the model molecular junction that we already discussed in Sec.\ \ref{Basics}. 
In contrast to the solid black lines, which has been obtained by the second-order 
rate equations (\ref{explicitsecondorderrates}), the effect of inelastic cotunneling and 
off-resonant pair creation processes is included in this result 
by employing the co-tunneling rates (\ref{cotrate}) and (\ref{cotrate2}). 
Significant differences between the two results appear only 
in the non-resonant transport regime, $e\Phi<2\overline{\epsilon}_{0}$. Due to the 
onset of inelastic co-tunneling processes (like the one depicted by Fig.\ \ref{IETSprocesses}b), 
the level of vibrational excitation increases at $e\Phi=\Omega$. It decreases again at higher bias voltages, 
when resonant deexcitation processes require less vibrational quanta. 
In the resonant transport regime, the influence of these processes 
appears to be negligible. 
The latter statement is corroborated by the dashed lines in Fig.\ \ref{vartemp} 
(which shows the level of vibrational excitation as a function of the electronic-vibrational 
coupling strength in the resonant transport regime, \emph{i.e.}\ $e\Phi=2(\overline{\epsilon}_{0}+3\Omega/2)$). 
They agree very well with the solid lines, which have been obtained by solving the 
second-order rate equations (\ref{explicitsecondorderrates}).

An analysis of the corresponding rate equations allows us to elucidate 
this point more clearly. 
In the limit $\lambda\rightarrow0$ and bias voltages $e\Phi\gtrsim2(\overline{\epsilon}_{0}+3\Omega/2)$, 
the rate equations (\ref{explicitcotunnelrates}) can be simplified to 
\begin{eqnarray}
0 &=& \Gamma_{\text{R}} \sigma^{\nu}_{1} - \Gamma_{\text{L}}  \sigma^{\nu}_{0} 
+ \Gamma_{\text{R}}  \lambda^{2} \left( (\nu+1) 
\sigma^{\nu+1}_{1} - (2\nu+1) \sigma^{\nu}_{1} + \nu \sigma^{\nu-1}_{1} \right) \\
&& - \lambda^{2} \left( (\nu+1) W_{\nu\rightarrow\nu+1} 
+ \nu W_{\nu\rightarrow\nu-1} \right) \sigma^{\nu}_{0} \nonumber\\
&& + \lambda^{2} \left( (\nu+1) W_{\nu\rightarrow\nu+1} \sigma^{\nu+1}_{0}
+ \nu W_{\nu\rightarrow\nu-1} \sigma^{\nu-1}_{0}  \right) \nonumber
\end{eqnarray}
and 
\begin{eqnarray}
0 &=& \Gamma_{\text{L}} \sigma^{\nu}_{0} - \Gamma_{\text{R}}  \sigma^{\nu}_{1} 
+  \Gamma_{\text{L}} \lambda^{2} \left( (\nu+1)  \sigma^{\nu+1}_{0} -  
(2\nu+1)\sigma^{\nu}_{0} + \nu  \sigma^{\nu-1}_{0} \right)   \\
&& - \lambda^{2} \left( (\nu+1) W_{\nu\rightarrow\nu-1} 
+ \nu W_{\nu\rightarrow\nu+1} \right) \sigma^{\nu}_{1} \nonumber\\
&& + \lambda^{2} \left( (\nu+1) W_{\nu\rightarrow\nu-1} \sigma^{\nu+1}_{1}
+ \nu W_{\nu\rightarrow\nu+1} \sigma^{\nu-1}_{1}  \right),  \nonumber
\end{eqnarray}
where we used $W_{\nu\rightarrow\nu\pm1}=\sum_{K,K'}\gamma_{K\rightarrow K',0\rightarrow0}^{\nu\rightarrow\nu\pm1}
=\sum_{K,K'}\gamma_{K\rightarrow K',1\rightarrow1}^{\nu\rightarrow\nu\mp1}$  
and, in contrast to our analysis in Sec.\ \ref{thermBroad}, we neglect thermal broadening, 
that is we evaluate each Fermi funtion by either 0 or 1 (including $f_{\text{L}}(\overline{\epsilon}_{0}+\Omega)$). 
Arguing again that 
$\Gamma_{\text{R}}\sigma_{1}^{\nu}-\Gamma_{\text{L}}\sigma_{0}^{\nu}=\mathcal{O}(\lambda^{2})$, we can further simplify 
these equations and obtain 
\begin{eqnarray}
 0 &=&  \left( 2 \Gamma_{\text{R}} + \frac{\Gamma_{\text{R}}}{\Gamma_{\text{L}}} W_{\nu\rightarrow\nu+1} + 
 W_{\nu\rightarrow\nu-1} \right) (\nu+1) \sigma_{1}^{\nu+1}  + \left( 2 \Gamma_{\text{R}} + \frac{\Gamma_{\text{R}}}{\Gamma_{\text{L}}} W_{\nu\rightarrow\nu-1} + 
W_{\nu\rightarrow\nu+1}   \right) \nu  \sigma^{\nu-1}_{1}   \nonumber\\
&&\hspace{-1cm} - \left( 2 \Gamma_{\text{R}} (2\nu+1) + \frac{\Gamma_{\text{R}}}{\Gamma_{\text{L}}}  \left( (\nu+1) W_{\nu\rightarrow\nu+1} 
+ \nu W_{\nu\rightarrow\nu-1} \right) 
+ \left( (\nu+1) W_{\nu\rightarrow\nu-1} 
+ \nu W_{\nu\rightarrow\nu+1} \right) \right) \sigma^{\nu}_{1}. \nonumber\\
\end{eqnarray}
The solution of this recurrence relation is 
\begin{eqnarray}
 \sigma_{1/0}^{\nu+1} = \sigma_{1/0}^{\nu}.  
\end{eqnarray}
which corresponds to an equal population of the vibrational levels and, therefore, to an 
infinite level of vibrational excitation. 
This result shows that, in the range of bias voltages considered, 
the ratio between excitation and deexcitation processes due to 
off-resonant pair creation and inelastic cotunneling processes 
is the same if broadening effects are explicitly excluded. Thus, the system 
exhibits a random walk through the ladder of all vibrational states, 
leading to the vibrational instability that has already been outlined in Refs.\ \cite{Semmelhack,Hartle2011}. 
We think this is an important finding, because broadening is typically understood 
in terms of these processes. An implementation of these processes 
in terms of the rates (\ref{cotrate}) and (\ref{cotrate2}) is often very useful. Our results show that 
this procedure has to be applied with care, as it can lead to qualitatively incorrect results for the 
level of vibrational excitation in the weak coupling limit $\lambda\rightarrow0$.

\subsection{Broadening due to coupling to a heat bath}
\label{heatbathsec}

\begin{figure}
\resizebox{\newwidth}{\newheight}{
\includegraphics{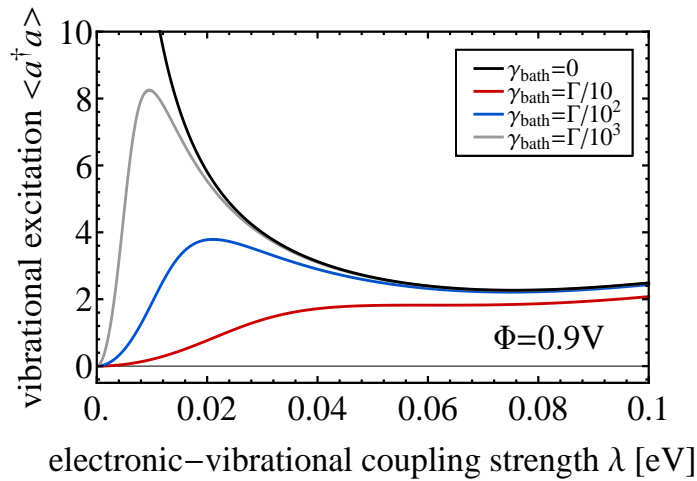}
}
\caption{\label{varbath} Level of vibrational excitation at $e\Phi=2(\overline{\epsilon}_{0}+3\Omega/2)$ 
as a function of the electronic-vibrational coupling strength $\lambda$ for different dissipation rates $\gamma_{\text{bath}}$.}
\end{figure}

In the previous sections, we were disregarding the effect of dissipation 
due to the coupling of the vibrational mode to a secondary bath of phonon modes (heat bath). 
We now include these effects via a non-zero dissipation rate $\gamma_{\text{bath}}$, solving the rate 
equations (\ref{explicitsecondorderrates}). 
Fig.\ \ref{varbath} depicts the corresponding level of vibrational excitation of our model molecular junction 
as a function of the electronic-vibrational coupling strength for different values of the 
dissipation rate $\gamma_{\text{bath}}$. 

The coupling to a heat bath appears to be particularly important 
for weak electronic-vibrational coupling strengths. In particular for $\lambda\rightarrow0$, 
it leads to an analytic behavior of this observable. The corresponding level of vibrational excitation 
tends towards the equilibrium value $(\text{exp}(\Omega/(k_{\text{B}}T))-1)^{-1}$. 
This behavior is independent of the value of $\gamma_{\text{bath}}$. 
The non-analytic behavior that we described before can therefore be considered to be an artifact 
of the model, where the dissipative effect of the environment on the local vibrational mode is neglected. 
Physical systems will always exhibit such coupling and, consequently, an analytic behavior 
of the level of vibrational excitation in the limit of vanishing 
electronic-vibrational coupling.

Our results also show that a negative slope of the level of vibrational excitation is obtained 
as long as the value of the dissipation rate $\gamma_{\text{bath}}$ is smaller 
than the excitation rate $\lambda^{2}\Gamma/\Omega^{2}$. A negative slope is not observed, 
once the coupling to the heat bath quenches the level of vibrational excitation also for strong 
electronic-vibrational coupling strengths $\lambda/\Omega\sim 1$.




\section{Conclusions}

We have studied the level of vibrational excitation in a model molecular junction 
with a single electronic state that is coupled to a single vibrational mode. 
We find that the level of vibrational excitation increases 
for a decreasing electronic-vibrational coupling strength, 
if deexcitation of the vibrational mode (local cooling) 
via electron-hole pair creation processes is suppressed by high enough bias voltages 
and if the coupling of the vibrational mode to a bath of secondary modes is not too strong. 
In the limit of vanishing electronic-vibrational coupling and without dissipation, 
the level of vibrational excitation 
is non-zero and exhibits non-analytic behavior. The corresponding value is finite and 
determined by the probability that electron-hole pair creation processes can occur. 
The latter is strongly influenced by the broadening of the junction's energy levels 
either due to thermal broadening or the hybridization with the Fermi sea of the electrodes. 
Including the coupling to a heat bath, the non-analytic behavior does not occur 
irrespective of the associated coupling strength, that is the non-analytic behavior 
is an artifact of neglecting the influence of the environment on the local 
vibrational mode.

Our findings are based on numerical and analytical results derived from Born-Markov theory. 
In addition, we considered transition rates corresponding to higher-order processes corresponding to, 
for example, off-resonant pair creation or inelastic cotunneling. While these processes 
can account for excitation and deexcitation mechanisms in the non-resonant transport regime, 
we have shown that a more rigorous treatment is necessary in the resonant regime. There, a  
treatment of these processes in terms of virtual tunneling processes, 
where broadening effects are explicitly excluded, leads to the same result as the one 
that is obtained from Born-Markov theory. 

At this point, we emphasize that a similar behavior of the level of vibrational excitation 
is found in the presence of a second higher-lying electronic state \cite{Hartle2011}. 
This is of particular interest in the present context, as the photon 
number in cQED systems has been measured in double quantum dot setups, 
using the spin-torque effect \cite{Viennot2014}. 
Future directions of research should therefore include the description of the 
level of vibrational excitation / photon number in the weak coupling limit of these 
more complex systems, including the effect of electron-electron interactions. 
Moreover, it may be interesting to develop schemes that allow to extract 
the photon number in single dot systems.

\section*{Acknowledgement}

We would like to thank T. Kontos for useful discussions. 
R.\ H.\ gratefully acknowledges financial support of the Alexander von Humboldt foundation 
via a Feodor Lynen research fellowship. 
We also thank the GWDG G\"ottingen for generous allocation of computing time. 
M.\ K.\ gratefully acknowledges support from the Professional Staff Congress -- 
City University of New York award \# 68193-00 46 
and thanks the hospitality of the Initiative for the Theoretical Sciences (ITS) - City University of New York 
Graduate Center and the Chemical Physics Theory Group of the University of Toronto where several interesting 
discussions took place during this work.

\end{document}